\documentclass{article}
\usepackage{PRIMEarxiv}

\usepackage[utf8]{inputenc} 
\usepackage[T1]{fontenc}    
\usepackage{hyperref}
\usepackage{url}            
\usepackage{booktabs} 
\usepackage{amsfonts}       
\usepackage{nicefrac}       
\usepackage{microtype}
\usepackage{lipsum}
\usepackage{fancyhdr}       
\usepackage{graphicx}
\graphicspath{{media/}}  

\usepackage{subfigure}
\usepackage{stfloats}
\usepackage{placeins}

\usepackage{wrapfig}
\usepackage{soul} 
\usepackage{color} 
\usepackage{makecell}
\usepackage{xcolor}         
\usepackage{multirow}
\usepackage{adjustbox}
\usepackage{wrapfig}
\usepackage{float}
\usepackage{bm}
\usepackage{amsmath}
\usepackage[noend]{algpseudocode}
\usepackage{algorithm}

\usepackage[capitalize,noabbrev]{cleveref}

\usepackage{enumitem}
\setlist{topsep=0pt, leftmargin=*}

\newcommand{\Name}{\texttt{VD-Gen} }

\newcommand{\revision}{}

\newcommand{\prepar}{\vspace{-4pt}}

\renewcommand{\algorithmicfunction}{\textbf{def}}

\title{3D Molecular Generation via Virtual Dynamics
}

\author{Shuqi Lu$^1$, Lin Yao$^1$, Xi Chen$^1$, Hang Zheng$^1$, Di He$^2$,  Guolin Ke$^1$ \\
$^1$DP Technology\\
$^2$Peking University\\
\{lusq, yaol, chenx, zhengh\}@dp.tech \\ 
 dihe@pku.edu.cn, kegl@dp.tech \\
}

\begin{document}
\maketitle

\begin{abstract}

Structure-based drug design, i.e., finding molecules with high affinities to the target protein pocket, is one of the most critical tasks in drug discovery. Traditional solutions, like virtual screening, require exhaustively searching on a large molecular database, which are inefficient and cannot return novel molecules beyond the database. The pocket-based 3D molecular generation model, i.e., directly generating a molecule with a 3D structure and binding position in the pocket, is a new promising way to address this issue. 
Herein, we propose VD-Gen, a novel pocket-based 3D molecular generation pipeline. 
VD-Gen consists of several carefully designed stages to generate fine-grained 3D molecules with binding positions in the pocket cavity end-to-end. Rather than directly generating or sampling atoms with 3D positions in the pocket like in early attempts, in VD-Gen, we first randomly initialize many virtual particles in the pocket; then iteratively move these virtual particles, making the distribution of virtual particles approximate the distribution of molecular atoms. After virtual particles are stabilized in 3D space, we extract a 3D molecule from them. Finally, we further refine atoms in the extracted molecule by iterative movement again, to get a high-quality 3D molecule, and predict a confidence score for it. Extensive experiment results on pocket-based molecular generation demonstrate that VD-Gen can generate novel 3D molecules to fill the target pocket cavity with high binding affinities, significantly outperforming previous baselines.  \looseness=-1

\end{abstract}

\vspace{-10pt}
\section{Introduction}
\label{Sec:intro}

Structure-based (pocket-based) drug design, i.e., finding a molecule to fill the cavity of the protein pocket with a high binding affinity~\cite{3DQSAR, desjarlais_using_1988, dewitte_smog_1997, dewitte_smog_1996}, is one of the most critical tasks in drug discovery. 
The most widely used method is virtual screening~\cite{walters1998virtual, shoichet_screening_2006, shoichet_virtual_2004}.
Virtual screening iteratively places molecules from a molecular database into the target pocket cavity and evaluates molecules with good binding based on rules such as energy estimation~\cite{de_ruiter_free_2011, chipot_free_2007, christ_basic_2010, michel_prediction_2010}. However, virtual screening is inefficient for the exhaustive search and is infeasible to generate new molecules that are not in the database.
Recently, molecular generative models have become a potential solution to address the problem as they could generate novel molecules in an efficient way. 
The early attempts focused on ligand-based molecular generation~\cite{DBLP:conf/icml/KusnerPH17, DBLP:conf/iclr/DaiTDSS18, winter2019efficient}, which trains models to learn the underlying distribution of the molecules in training data and generate similar molecules. 
However, those methods did not consider conditional information, such as the shape of the pocket. Later, more efforts were paid to studying how to generate molecules conditioned on the information of protein pockets. 
Some pocket-based generative models simply generate molecules in the form of SMILES or graphs~\cite{skalic_target_2019, xu_novo_2021}, without considering the 3D geometric position of the molecule and pocket, which is closely related to binding affinity.
\looseness=-1

Given the 3D structure of a pocket, the ultimate goal of the task is to generate 3D molecules which contain a set of atoms, each with an atom type and the corresponding 3D position. 
Previous works can be roughly categorized into the following two classes.
1) 3D density grid generation \cite{ragoza_generating_2022}, in which pockets and molecules are converted to 3D density grids with coarse-grained positions, and then a generative model is used to predict the density at each grid.
Since the model can only generate grid-level positions, these approaches cannot obtain high-quality 3D molecules.
2) Auto-regressive 3D generation~\cite{sbdd,  liu_generating_2022, peng_pocket2mol_2022}, in which atoms (with a 3D position and an atom type) are generated one by one. 
However, since it is hard to define which atoms should be generated first during training, these models usually achieve inferior performance. \looseness=-1

 \begin{figure*}[!tbp]%
 \vspace{-4pt}
 \centering
\includegraphics[width=0.95\linewidth]{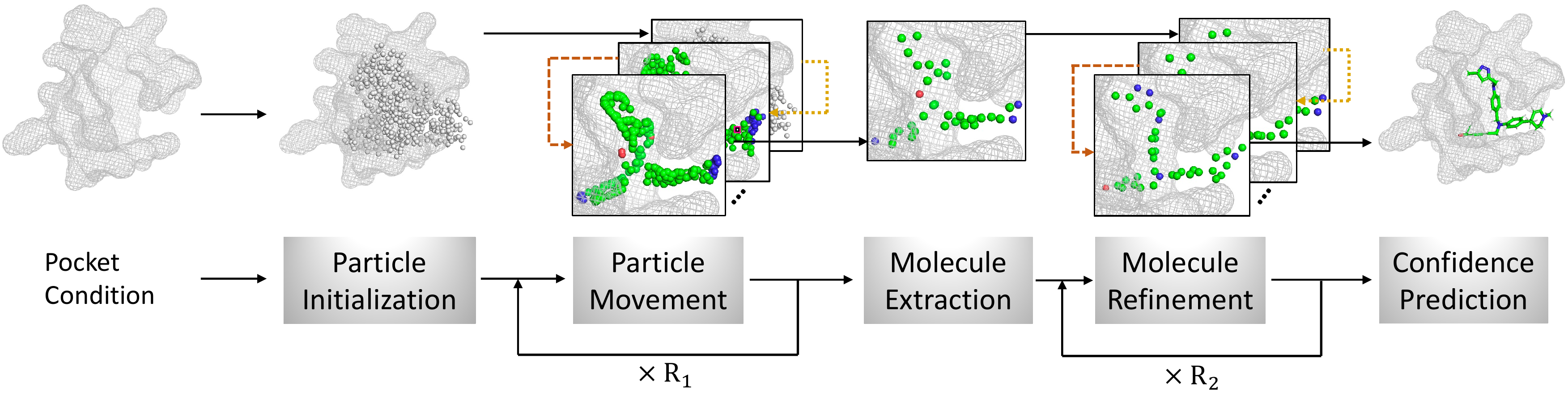}
\vspace{-10pt}
\caption{The pipeline of \texttt{VD-Gen}. Given a pocket as the condition, it first initializes the Virtual Particles (VPs), and iteratively moves them, to approximate the atom distribution. Then, a 3D molecule is extracted from the VPs after movement. Next, the atoms in the 3D molecule are refined by iterative movement again. Finally, a confidence score for the generated 3D molecule will be predicted.
} \label{fig:model}
\end{figure*}%




In this paper, we proposed a new method \Name, which can generate high-quality 3D molecules efficiently. The key idea of \Name is using a distribution of Virtual Particles (VPs) in 3D space to represent the distribution of molecular atoms in 3D space. Specifically, as shown in Fig.~\ref{fig:model}, \Name pipeline contains 5 stages to generate 3D molecules end-to-end.
1) Given a protein pocket, \Name first initializes multiple VPs with types and positions.
2) It then iteratively moves the VPs until equilibrium. Ideally, the distribution of equilibrious VPs will be close to the distribution of molecular atoms.
3) A 3D molecule is then extracted from the equilibrious VPs.
4) It then refines the atoms in the extracted molecule, by iterative movement again.
5) A confidence score for the generated 3D molecule will be predicted. 
With the above pipeline, \Name can generate high-quality 3D molecules non-auto-regressively, and addresses the issues in the previous works. Compared with 3D grid based generative models, \Name can generate high-quality 3D molecules with fine-grained coordinates. Compared with auto-regressive based generative models, \Name can efficiently generate all atoms at once, and thus the performance is better and not related to the generation order.  \looseness=-1

We conduct extensive experiments with multiple evaluation metrics, such as Vina~\cite{vina}, MM-PBSA~\cite{yang_wang_zheng_2022}, 3D Similarity~\cite{ligsift}, to benchmark \texttt{VD-Gen} thoroughly. The experimental results show that our model can generate diverse drug-like molecules with high binding affinities in 3D space with good binding poses, significantly outperforming all baselines. Ablation studies, case studies, and visualizations are designed to further demonstrate the effectiveness of \texttt{VD-Gen}. In addition, \Name pipeline can be easily extended to pocket-based 3D molecular optimization, achieving superior performance as well.

\vspace{-8pt}
\section{Method} \label{Sec:method}
\vspace{-5pt}

The goal of pocket-based 3D molecular generation is to learn $\mathbf{M} = h(\mathbf{P} ; \bm{\theta})$, where $h(\cdot ; \bm{\theta})$ is a model with parameter $\bm{\theta}$, $\mathbf{P} = \{(\bm{x}_i^p, \bm{y}_i^p)  \}^u_{i=1}$ is the set of $u$ atoms in the pocket,  $\bm{x}_i^p \in \mathbb{R}^t$ and $\bm{y}_i^p \in \mathbb{R}^3$ are the $i$-th pocket atom's type (one-hot) and coordinate, respectively, $t$ is the number of atom types, and $\mathbf{M} = \{(\bm{x}_i, \bm{y}_i)  \}^m_{i=1}$ is the set of $m$ atoms of the generated molecule.

Rather than generating $\mathbf{M}$ directly, \Name first models the atom distribution given the pocket (i.e., conditional on $\bm{P}$) in 3D space, then extracts the molecules from the distribution. 
To learn an atom distribution towards the ground-truth atom positions of each atom type, we adopt a learning strategy similar to Molecular Dynamics. In particular, we introduce Virtual Particles (VPs), which have types and 3D coordinates like atoms. A set of VPs is randomly allocated in a predicted region in the pocket cavity,  acting as a distribution. Then, the VPs are gradually moved based on a learnable dynamics, approximating the ground truth molecular atoms. We call this iterative process \texttt{Virtual Dynamics}.

Please note \Name models the distribution of atoms, not the distribution of molecules. Thus, we cannot directly sample molecules from the atom distribution. To extract the molecules from the atom distribution, we design a filter-then-merge method to extract atoms from the clusters of VPs, and then continued to refine the extracted atoms by \texttt{Virtual Dynamics} again. Besides, considering the practice usage, \Name also predicts confidence scores for the generated 3D molecules. We describe the overall pipeline of \Name in the next subsection, and we also summarize the overall inference pipeline in the Alg.~\ref{alg:overall_infer}.

\prepar
\subsection{VD-Gen Pipeline}


\prepar
\paragraph{Particle Initialization} 
The goal of this stage is to initialize VPs inside the protein pocket. First, we use a neural network model to predict the number of molecular atoms based on pocket atoms, denoted as $\bar{m} = h_{an} ( \mathbf{P} ; \bm{\theta}_{an})$, where $h_{an}$ is a neural model with learnable parameter $\bm{\theta}_{an}$. And the number of VPs is set as $n = \bar{m}k_{vp}$, where $k_{vp} \in \mathbb{N}$ is a hyper-parameter. 
Then, we use another model to determine where to allocate the VPs. Given the atoms of a pocket $\mathbf{P}$, we build a 3D grid cubic with binary voxel values ("1" means the grid has pocket atoms), and use a 3D U-Net model~\cite{unet} to predict the grids that may contain the molecular atoms. Formally, we denote this process as $
\mathbf{C}_m = h_{pi} ( \mathbf{C}_p  ; \bm{\theta}_{pi})
$, where $h_{pi}$ is 3D U-Net model with learnable parameter $\bm{\theta}_{pi}$, $\mathbf{C}_p \in \{0, 1\}^{l_1 \times l_2 \times l_3}$ is the gridded 3D cubic (with size $l_1 \times l_2 \times l_3$) of pocket atoms, and $\mathbf{C}_m \in \{0, 1\}^{l_1 \times l_2 \times l_3}$ is the predicted cubic, in which the grids with voxel value 1 may contain the molecular atoms. Finally, we randomly and uniformly distributed $n$ VPs in the grids with voxel value 1 in $\mathbf{C}_m$. And we use $\mathbf{V}_0$ to denote the initialized VPs.

\prepar
\paragraph{Particle Movement} 

Given $\mathbf{V}_0$, we then update the distribution of them by moving them in 3D space, to approximate the ground-truth atom distribution.
Similar to Molecular Dynamics, the movement in this stage is iterative. 
In particular, at each iteration, the model will take the VPs' positions and types from the previous iteration as inputs, and output the new positions and types for them. 
This process could be denoted as $\mathbf{V}_{r+1} = f_{pm}(\mathbf{V}_{r}, \mathbf{P}; \bm{\theta}_{pm})$, $\mathbf{V}_r = \{(\bm{x}_i^r, \bm{y}_i^r)  \}^n_{i=1}$ is the set of $n$ VPs that are predicted at the $r$-th iteration, $f_{pm}$ is a SE(3) model that can take 3D coordinates as inputs, and $\bm{\theta}_{pm}$ is the learnable parameters. And we use $R_1$ to denote the number of iterations in \emph{Particle Movement} stage.

\prepar
\paragraph{Molecule Extraction} Since $\mathbf{V}_{R_1}$ produced by \emph{Particle Movement} stage is an approximation of atom distribution, not the molecule distribution, we cannot directly sample molecules from it. Therefore, in this stage, we design a method to extract molecules from $\mathbf{V}_{R_1}$. Formally, in this step, the model can be denoted as $\mathbf{W}_0 = h_{me} (\mathbf{V}_{R_1}, \mathbf{P}; \bm{\theta}_{me})$, where $\bm{\theta}_{me}$ is learnable parameters, $\mathbf{W}_0 = \{(\hat{\bm{x}}_i^0, \hat{\bm{y}}_i^0)\}_{i=1}^m$ is the set of $m$ atoms of the extracted 3D molecule. The model reduces $n$ VPs to $m$ atoms by two steps, filter and merge. First, as some VPs may fail to approach their target positions, we want to filter out them. The errors (distances between the VPs and their target positions) for VPs are predicted, and the VPs with errors larger than $\zeta$, a hyper-parameter, will be filtered out.
Second, we want to merge the remaining VPs into atoms. The model will predict a merging probability of a pair of two VPs. With the predicted pair-wise merging probability matrix, we can then use a threshold to get a binary merging matrix and merge VPs into clusters according to the matrix. However, it is hard to decide a threshold. Notice that we had predicted the number of atoms $\bar{m}$ in \emph{Particle Initialization}. We can find a merging threshold by binary search, making the number of clusters approximate $\bar{m}$. \looseness=-1

Then there will be several (ideally $\bar{m}$) merged clusters, and we denote $\bm{w}_i$ as the set of the indices of $i$-th cluster's VPs. Then, to initialize $\mathbf{W}_0$, we use $\hat{\bm{x}}_i^0 = \text{Uniform} (\{\bm{x}^{R_1}_j | j \in \bm{w}_i \})$ and $\hat{\bm{y}}_i^0 = \text{Mean} (\{\bm{y}^{R_1}_j  | j \in \bm{w}_i \})$, to sample an atom type and get an average coordinate respectively.

\prepar
\paragraph{Molecule Refinement} Given the extracted molecule from \emph{Molecule Extraction}, we further refine its atoms $\mathbf{W}_0$, to get a high-quality 3D molecule with fine-grained 3D coordinates.
Similar to \emph{Particle Movement}, we move the atom positions iteratively, but with different model parameters. Formally, this stage can be denoted as $\mathbf{W}_{r+1} = f_{mr}(\mathbf{W}_{r}, \mathbf{P}; \bm{\theta}_{mr})$, where $f_{mr}$ is SE(3) model with
learnable parameter $\bm{\theta}_{mr}$, and $\mathbf{W}_r = \{(\hat{\bm{x}}_i^r, \hat{\bm{y}}_i^r)\}_{i=1}^m$ is the set of $m$ atoms at the $r$-th iteration. 
And we use $R_2$ to denote the total iterations in \emph{Molecule Refinement} stage. 

\prepar
\paragraph{Confidence Prediction} 
In real-world tasks, we usually need to generate more than one molecule and select the top ones among them. Therefore, a confidence predictor is needed to select or rank the molecules according to binding affinities. Although we can use computational simulations or wet experiments to examine the generated molecules, they are too costly, especially for a large number of molecules. To further improve the usability of \Name and reduce the extra cost of selecting good molecules, a confidence score for each generated 3D molecule will be predicted in this stage. \looseness=-1

\prepar
\subsection{VD-Gen Training Strategies} \label{sec:train}


\prepar
\paragraph{Training of Particle Initialization} 
Given the protein-ligand complex data, we can easily train the 2 neural models of this stage. For the atom number prediction model $h_{an}$, we can directly get the training label, i.e., the ground-truth number of atoms, from the ligand molecules in the complex data. To stabilize the training, we bucket the number of atoms into one-hot bins, converting the regression task to a classification task: \looseness=
-1
\begin{equation}
\small
\mathcal{L}_{an} = \text{NLL}(\bar{\bm{o}}, \bm{o}),
\end{equation}
where NLL is the negative log likelihood loss function, $\bm{o}$ is the one-hot vector of the bucketed atom number, and $\bar{\bm{o}}$ is the predicted vector from model $h_{an}$.

For 3D U-Net $h_{pi}$, we can get the training ground-truth label $\mathbf{C}_m^g$ for $\mathbf{C}_m$, i.e., the gridded cubic of molecular atoms, from the ligand molecules in the complex data. And the training objective function is the grid-wise binary classification. Besides, we additional use a focal loss~\cite{lin2017focal} to relieve the unbalanced classification problem:
\begin{equation}
\small
\mathcal{L}_{pi} = \frac{1}{l_1 \times l_2 \times l_3} \sum_{i=1}^{l_1} \sum_{j=1}^{l_2} \sum_{k=1}^{l_3} \text{FL}(\bar{\bm{u}}_{i,j,k}, \bm{u}_{i,j,k})
\end{equation}
where FL is the focal loss function, $\bm{u}_{i,j,k}$ is the one-hot vector of ground-truth voxel in $\mathbf{C}_m^g$, $\bar{\bm{u}}_{i,j,k}$ is the predicted vector from model $h_{pi}$. 

\prepar
\paragraph{Training of Particle Movement} The goal of model $f_{pm}$ is to move the VPs to the positions of ligand molecular atoms, so that the distribution of VPs can approximate the distribution of molecular atoms. To achieve this, we can directly assign a real atom as the training target for each VP.
Formally, given the ground-truth atoms $\mathbf{G} = \{(\bm{x}_i^g, \bm{y}_i^g)  \}^m_{i=1}$ and the initialized VPs $\mathbf{V}_{0}$, there are $n^m$ possible assignments. Following the principle of least action~\cite{feynman2017character}, the assignment with minimal moving distance is favored. That is to optimize $\text{Min} \sum_{i=1}^n \lVert \bm{y}_i^0 - {\bm{y}}^g_{a_i} \rVert_2$, where $\bm{y}_i^0$ is the initial position and $a_i \in \mathbb{N}$ is the assigned target for $i$-th VP. This optimization problem is easy to solve: for $i$-th VP, assign its nearest real atom as the training target, i.e., $a_i=\text{arg min}_{j=1}^{m} \lVert \bm{y}_i^0 - {\bm{y}}^g_j \rVert_2$.

Given the assigned targets $a_i$, we use the following losses for the training. First, a negative log likelihood loss is used for the VPs' types. Second, a clip L2 loss is used for the VPs' 3D coordinates. Third, two L1 losses are used for the VP-VP pair distances and VP-pocket pair distances, respectively. Finally, a regularization loss is used to limit the moving distances between two adjacent iterations. Combined above, the final training loss function at the $r$-th iteration could be denoted as \looseness=-1
\begin{equation}
\small
\begin{aligned}
\label{equ:atom_coord_loss}
\mathcal{L}_{pm} &= \frac{1}{n} \sum_{i=1}^{n} \left( \text{NLL}(\bar{\bm{x}}_i^r, \bm{x}^g_{a_i}) 
 + {\color{black}\text{clip}}(||{\bm{y}}_i^r - \bm{y}_{a_i}^g||_2, \tau) \right.\\ &\left.  +  \frac{1}{n} \sum_{j=1}^{n}||{\bm{d}}_{ij}^r - \bm{d}_{a_i,a_j}^g ||_1  + \frac{1}{u} \sum_{j=1}^{u}||{\bm{c}}_{ij}^r - \bm{c}_{a_i,j}^g ||_1  \right.\\ &\left. 
 +  {\color{black}\text{max}}(||{\bm{y}}_i^r - {\bm{y}}_i^{r-1}||_2-\delta,0) \right),
\end{aligned}
\end{equation}
where $\bar{\bm{x}}_i^r$ is the predicted vector of atom types of $i$-th VP, $\bm{x}_{a_i}^g$ is the ground-truth atom types of $i$-th VP, $\bm{y}_i^r$ ($\bm{y}_{a_i}^g$) is the predicted (ground-truth) coordinates of $i$-th VP, $\tau$ is the the clip value for coordinate loss, ${\bm{d}}_{ij}^r$ ($\bm{d}_{a_i,a_j}^g$) is the predicted (ground-truth) distance of the $i$-th and $j$-th VP pair, ${\bm{c}}_{ij}^r$ ($\bm{c}_{a_i,j}^g$) is the predicted (ground-truth) distance of the $i$-th VP and the $j$-th pocket atom, and $\delta$ is the threshold for moving regularization.

However, training the model with multiple iterations is not efficient in both speed and memory consumption. To reduce the training cost, we adopt the stochastic iteration in AlphaFold2~\cite{jumper_highly_2021}. In particular, during training, the iteration $r$ is uniformly sampled between $1$ and $R$, where $R$ is the max iteration ($R=R_1$ in this stage). Then, the model is run on the forward-only mode in the first $r-1$ iterations, without loss calculation and gradient backward. Finally, the gradient and backward are enabled at the $r$-th iteration. During inference, the sampling on iterations is not used. The above algorithm is shown in Alg.~\ref{alg:iterm}.


\prepar
\paragraph{Training of Molecule Extraction}
There are two training tasks in this stage. The first is to predict the errors (distances between the VPs and their target positions). To stabilize the training, we bucket the errors into one-hot bins, converting it to a classification task:
\begin{equation}
\small
\mathcal{L}_{error\_pred} = \frac{1}{n} \sum_{i=1}^n\text{NLL}( \bar{\bm{s}}_i, \bm{s}_i),
\end{equation}
where $\bm{s}_i$ is the one-hot vector of the bucketed target error bin,  
and $\bar{\bm{s}}_i$ is the predicted probability vector. 

The second is to predict which VP pairs should be merged. Ideally, the VPs with the same target atom should be merged, thus training label for 
a VP pair with the same atom target is set to "true". When there are $n$ VPs and $m$ real atoms, the ratio of "true" class is about $\frac{m \times (n / m)^2}{n^2} = \frac{1}{m}$. As $m$ ranges from dozens to hundreds, the binary classification task here is very unbalanced. Thus, we introduce a focal loss~\cite{lin2017focal} to balance the classes.
\begin{equation}
\small
\mathcal{L}_{merge} =  \frac{1}{l^2} \sum^l_{i=1}\sum^l_{j=1} FL (\bar{\bm{r}}_{ij}, \bm{r}_{ij}),
\end{equation}
where $l$ is the number of virtual particles after filtering, $\bm{r}_{ij}$ is the target merging type, $\bar{\bm{r}}_{ij}$ is the predicted probability of merging type.

\prepar
\paragraph{Training of Molecule Refinement}
This training is very similar to \emph{Particle Movement}, except the training target is different. In particular, the training target for the $i$-th atom is the most frequent target atom in the cluster $\bm{w}_i$, not its nearest atom. Formally, for the $i$-th atom, its target atom is denoted as $b_i = \text{most\_frequent}(\{a_j | j \in \bm{w}_i  \})$. 

\prepar
\paragraph{Training of Confidence Prediction}

We explicitly train a task to learn the confidence scores for the generated molecules. In particular, following AlphaFold~\cite{jumper_highly_2021}, we compute the LDDT score~\cite{mariani2013lddt} of the generated molecule and ground-truth molecule, and a model is used to predict the LDDT score. Also, we bucket the LDDT score to one-hot bins, converting it to a classification task. 
\begin{equation}
\mathcal{L}_{confidence} = \frac{1}{n} \sum_{i=1}^n \text{NLL}(\bar{\bm{e}}_i, \bm{e}_i),
\end{equation}
where $\bar{\bm{e}}_i$ is the predicted LDDT probability distribution, and $\bm{e}_i$ is the one-hot vector of the bucketed LDDT bins. 

\prepar
\paragraph{SE(3) Model} 
Both $f_{pm}$ and $f_{mr}$ require to be SE(3) models that can take 3D coordinates as inputs, and outputs new 3D coordinates.  We mainly follow the design of the efficient SE(3)-equivariance Transformer proposed in Uni-Mol~\cite{zhou2022uni} and Graphormer-3D~\cite{shi2022benchmarking}. However, they did not consider the interaction between pocket and molecule. Therefore, we extend the model with an additional pocket encoder, and add a particle-pocket attention to capture the interactions between pocket atoms and VPs. Since this paper focuses on the pocket-based 3D molecular generation, not the SE(3) models, we leave the details of the designed SE(3) model in Appendix~\ref{app:backbone}.  \looseness=-1

\begin{algorithm}[t]

\caption{Iterative Movement}\label{alg:iterm}
\begin{algorithmic}[1]
\small
\revision
\Require {$R$: max iterations, $\mathbf{P}$: pocket atoms, $\mathbf{V}_0$: random initialized VPs, $f(\cdot ; \bm{\theta})$: SE(3) model with parameters $\bm{\theta}$}
\State $r \gets$ uniform(1, $R$) if \sethlcolor{yellow}\hl{training} else $R$ \Comment{Sampling is only enabled at training}
\State \sethlcolor{yellow}\hl{disable\_gradient()} \Comment{Disable gradient calculation globally}
\For{$k \in [1,...,r - 1)$} 
\State $\mathbf{V}_k  \gets f(\mathbf{V}_{k-1}, \mathbf{P}; \bm{\theta} )$  
\Comment{update without gradients}
\EndFor
\State \sethlcolor{yellow}\hl{enable\_gradient()} \Comment{Enable gradient calculation globally}
\State $\mathbf{V}_{r}  \gets f(\mathbf{V}_{r-1}, \mathbf{P}; \bm{\theta} $) \Comment{update with gradients} \\
\Return  $\mathbf{V}_{r}$ \Comment{Return the positions and types of particles}
\end{algorithmic}
\end{algorithm}




\prepar
\subsection{Extending VD-Gen to 3D Molecular Optimization}

Molecular optimization is also an important task in real-world drug design. In molecular optimization, rather than generating from scratch, the goal is to replace a part of the given molecule, like a fragment, and to get a molecule with better binding affinity. Here, we extend \texttt{VD-Gen} to the pocket-based 3D molecular optimization. In particular, as illustrated in Fig.~\ref{fig:model_optim}, we first randomly remove a fragment of the given molecule, and the model is learned to generate it, with the pocket and the remaining atoms in the molecule as conditions. In this way, although it is not trained to optimize molecules directly, the model learns how to remove-then-fill a fragment of a molecule, and thus could be used in molecular optimization tasks. The benchmark results of molecular optimization are left to Appendix~\ref{Appendix:model_optim}.

\section{Experiments}
\label{sec:experiments}


\subsection{Settings}

\prepar
\paragraph{Evaluation metrics} There is not a golden metric to evaluate the generated molecules, so we use multiple metrics to have a comprehensive evaluation. 
1) \emph{3D Similarity}. As the pocket-based 3D generation models are trained by the 3D structures of the pockets and molecules, the most direct metric to examine the models' generative ability is to evaluate the 3D similarity between the generated molecule and the ground-truth one. Here we use LIGSIFT~\cite{ligsift} to calculate the overlapping ratio in 3D space between two molecules.
2) \emph{Vina}. Docking scores, like \text{Vina}~\cite{vina}, are widely used in previous pocket-based generation works, for they are easy to compute. To be consistent with previous works, we also use \textbf{Vina} as a metric. However, previous works usually relied on Vina's re-docking, in which the molecular conformation and binding pose may be largely changed by docking tools. Thus, to directly evaluate the 3D molecules generated by model, we add an additional \textbf{Vina*} score that does not use re-docking. 
3) \emph{MM-PBSA}. Although docking scores are easy and fast to compute, they are proposed to recall the possible hits in the large-scale virtual screening, not for ranking. Thus, docking scores are not good metrics to compare the binding affinities for different models \cite{cheng2009comparative}, and we further use the slower but more accurate MM-PBSA (Molecular Mechanics Poisson–Boltzmann Surface Area) \cite{mmpbsa} as a metric. Based on \text{MM-PBSA}, we add two additional metrics. \textbf{MM-PBSA B.T.} (MM-PBSA Better than Target), which computes the percentage of generated molecules with better MM-PBSA scores than ground-truth. \textbf{MM-PBSA Rank}, which computes the average rankings of different models among different complexes. Due to MM-PBSA scores varying largely in different complexes, \text{MM-PBSA Rank} can better compare different models. 
The details of the above metrics are described in Appendix~\ref{Appendix:metric}. \looseness=-1

\prepar
\paragraph{Data} 
We use the same training data as in previous works\cite{sbdd, peng_pocket2mol_2022}, 
CrossDocked data \cite{francoeur2020three}, to train \Name. The training set contains 100,000 protein-ligand pairs.
For the test set, we use 100 protein-ligand complex crystal structures from \cite{yang_wang_zheng_2022}, on which MM-PBSA was validated to be effective. 
To avoid leakage, we also remove the training data's complexes whose protein sequences are similar to the ones in the test set. In particular, two protein sequences are identified as similar if their e-value from BLAST \cite{blast} search results is larger than 0.4. \looseness=-1

\prepar
\paragraph{Training}  



There are 3 models in the \Name, a 3D U-Net used to predict where to initialize VPs, two SE(3) models (details in Sec.~\ref{app:backbone}) used for \emph{Particle Movement} and \emph{Molecule Refinement}, respectively. These two SE(3) models share the same pocket encoder.  The atom number is predicted by a head at the pocket encoder. The two prediction tasks in \emph{Molecule Extraction} are predicted based on the SE(3) model used in \emph{Particle Movement}, by two additional heads. The confidence score is predicted by a head at \emph{Molecule Refinement}'s SE(3) model. The number of total parameters is about 144M. 

We first train the 3D U-Net model standalone, and it costs about 20 hours by 8 NVIDIA A100 GPUs. Then, we freeze the parameters of the 3D U-Net model, and train the whole \Name pipeline end-to-end. The training of the whole pipeline costs about 15 hours by 8 NVIDIA A100 GPUs.


We leave the detailed hyper-parameters used in training to Appendix~\ref{sec:app:args}.

\begin{table*}[t]
\vspace{-8pt}
  \centering
  %
  \caption{Performance on pocket-based 3D molecular generation.}
  \vskip 0.15in
    \begin{tabular}{c|c|cc|rrrr}
    \toprule
    \multirow{2}{*}{Model} & 3D Sim($\uparrow$) & Vina($\downarrow$) & Vina*($\downarrow$) & MM-PBSA($\downarrow$)   & MM-PBSA-   & MM-PBSA-  \\
    & & & &   &  Rank($\downarrow$) & B.T.(\%$\uparrow$)  \\
    \midrule
    LiGAN~\cite{ragoza_generating_2022}  & 0.356  &  -6.724 &  -5.372 & -18.462 & 2.59 & 0.3   \\
    3DSBDD~\cite{sbdd}  & 0.365  &  -8.662 &  -7.227 & -30.560  & 2.31  &  2.29  \\
    GraphBP~\cite{liu_generating_2022} & 0.333 & -8.710 & -3.689  & -5.579 & 4.01  & 0   \\
    Pocket2Mol~\cite{peng_pocket2mol_2022} & 0.352 & -8.332 &  -6.525 & -8.226   & 3.53  & 0 \\

    \texttt{VD-Gen} & \textbf{0.422} & \textbf{-8.998} & \textbf{-7.397} & \textbf{-50.749} & \textbf{1.16} & \textbf{11.7} \\
    
    \bottomrule
    \end{tabular}
  \label{tab:binding_overall}
\vskip -0.1in
\end{table*}


\begin{figure*}[ht]
\centering
\subfigure[Number of VPs\label{fig:ablation_vp_num}]{
\begin{minipage}[t]{0.23\linewidth}
\centering
\includegraphics[width=4.1cm]{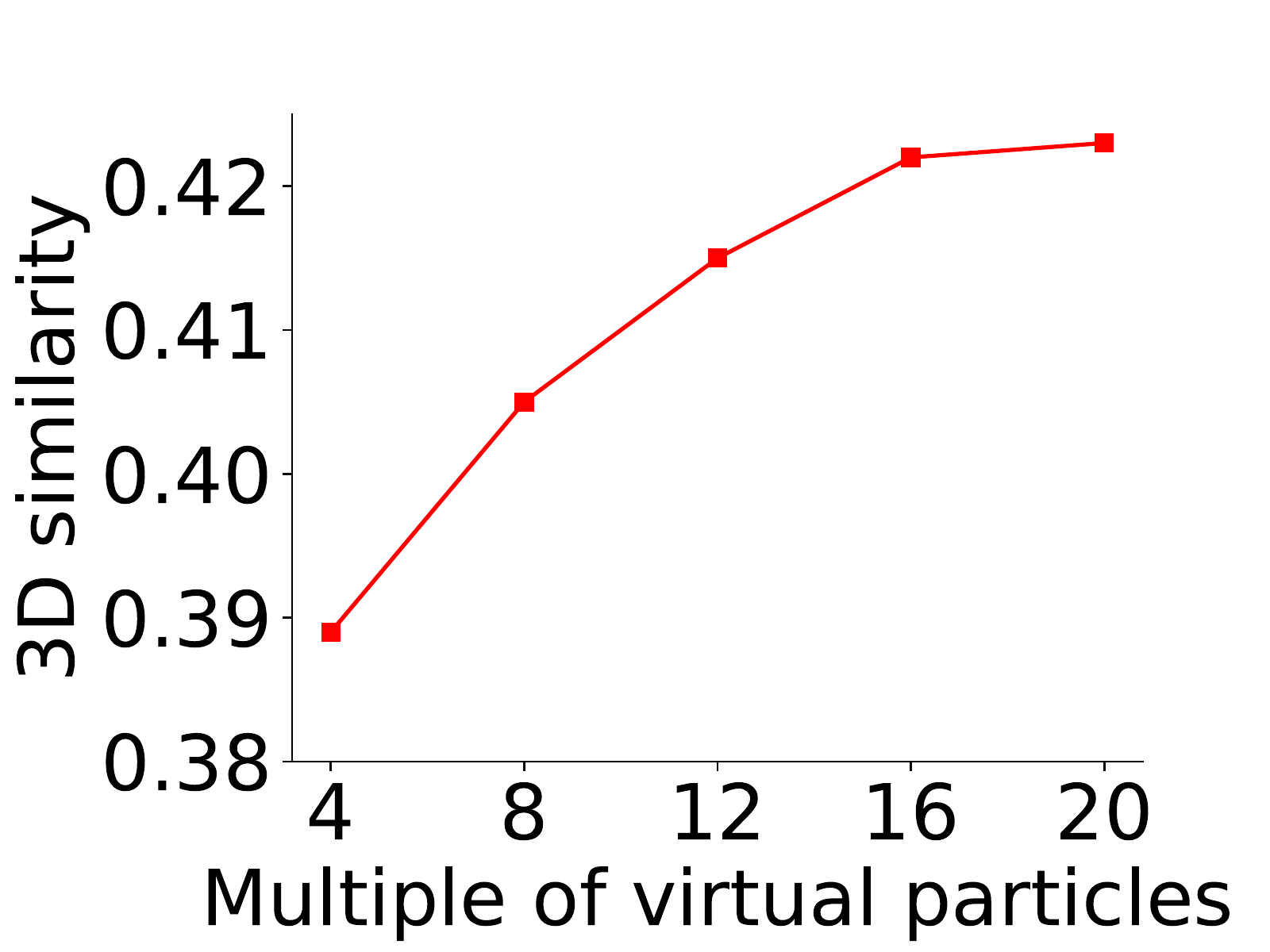}
\end{minipage}
}
\subfigure[Particle Movement \label{fig:ablation_equ_iteration}]{
\begin{minipage}[t]{0.23\linewidth}
\centering
\includegraphics[width=4.1cm]{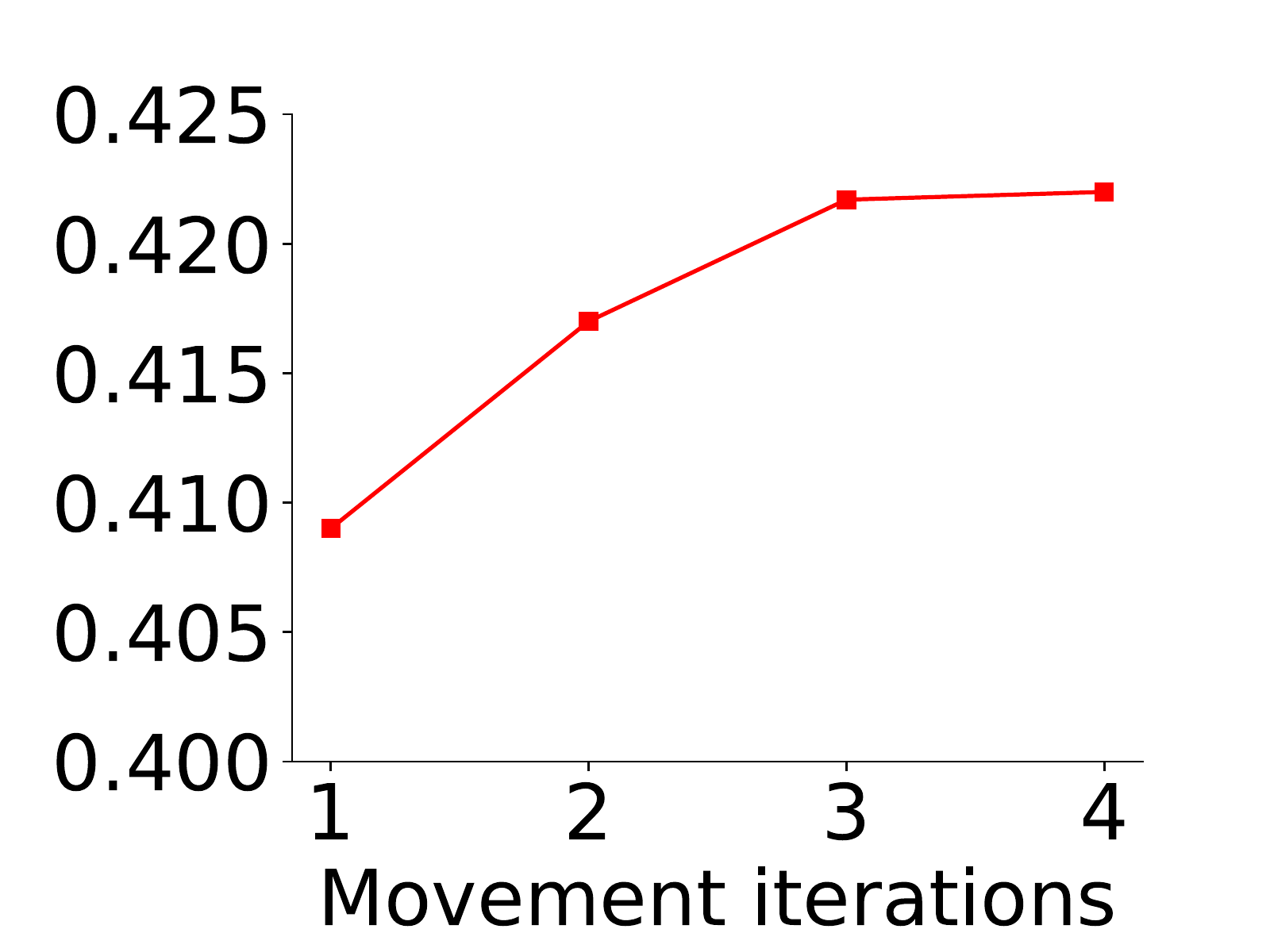}
\end{minipage}
}
\subfigure[Molecule Refinement\label{fig:ablation_refine_iteration}]{
\begin{minipage}[t]{0.23\linewidth}
\centering
\includegraphics[width=4.1cm]{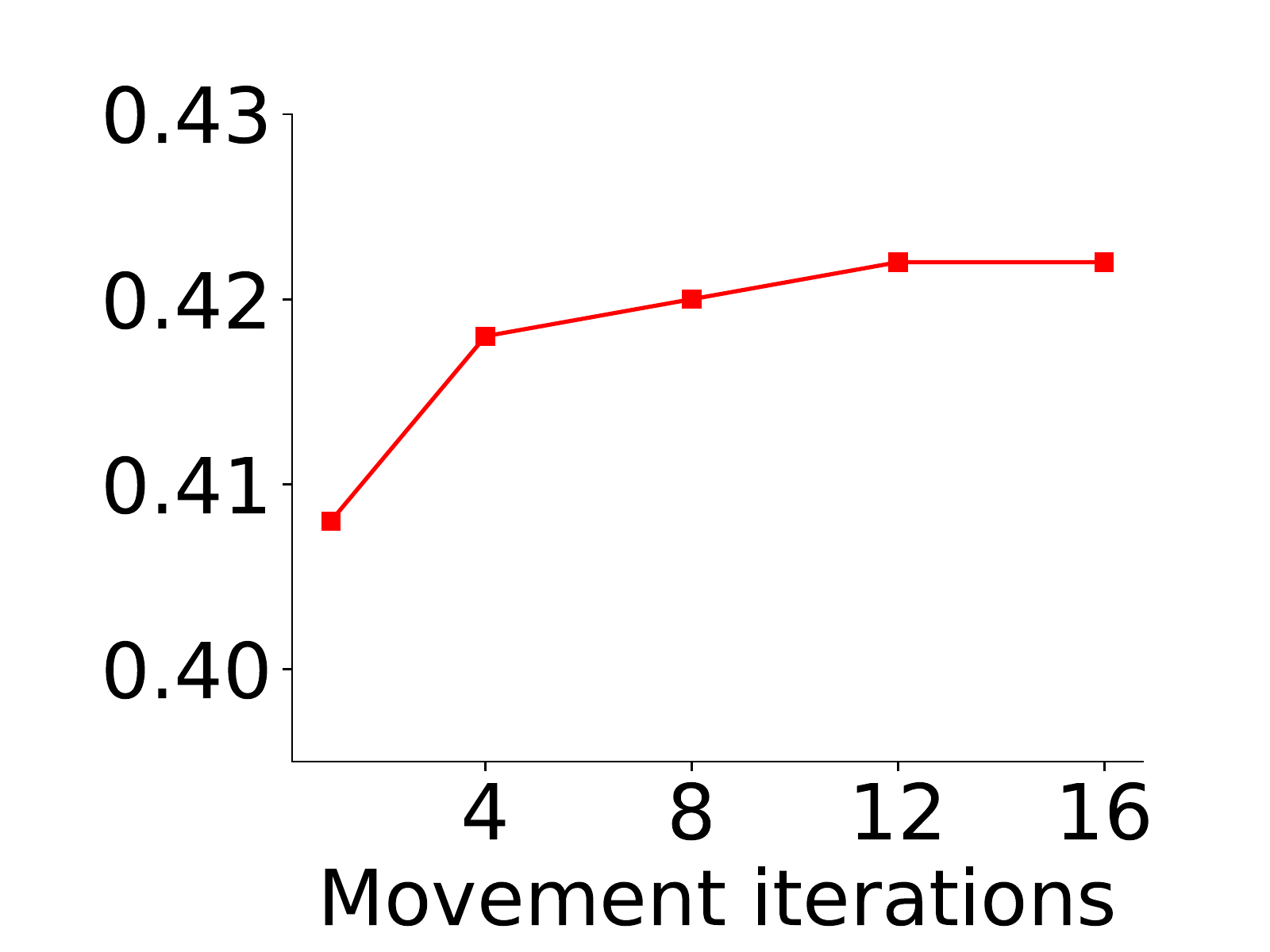}
\end{minipage}
}
\subfigure[Confidence Prediction\label{fig:ablation_relevance_3d}]{
\begin{minipage}[t]{0.23\linewidth}
\centering
\includegraphics[width=4.1cm]{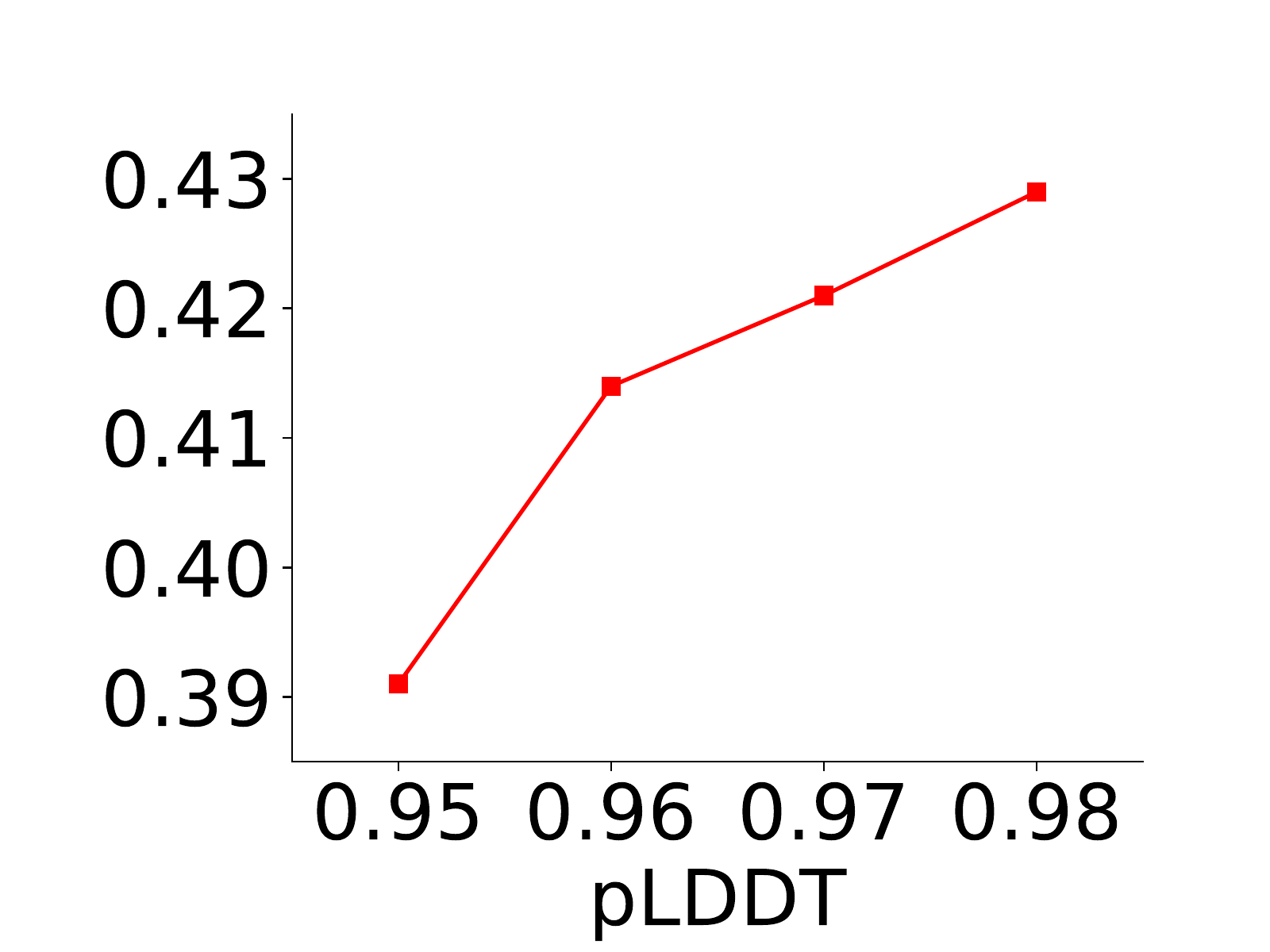}
\end{minipage}
}
\vspace{-10pt}
\caption{Ablation studies for \texttt{VD-Gen} pipeline.}
\label{fig:ablation_result}
\end{figure*}


\prepar
\subsection{Molecule Generation Performance}

\prepar
\paragraph{Baselines}
We compare \texttt{VD-Gen} with several previous 3D pocket-base molecular generation models: the 3D density grid generative model LiGAN~\cite{ragoza_generating_2022}, and the auto-regressive 3D generative models GraphBP~\cite{liu_generating_2022}, 3DSBDD~\cite{sbdd}, and Pocket2Mol~\cite{peng_pocket2mol_2022}. For all models, we generate 500 molecules for each pocket, and then select 100 from them for evaluation. For 3DSBDD and Pocket2Mol, beam search is used and the top 100 molecules are selected. For \texttt{VD-Gen}, the selection is based on the confidence score. For LiGAN and GraphBP, random 100 molecules are selected due to they did not implement beam search.

\prepar
\paragraph{Results} As we pay more attention to the generated molecules with high binding affinities, we report the top 5-th percentile result for Vina, Vina*, and MM-PBSA. MM-PBSA-Rank is calculated based on the top 5-th percentile MM-PBSA result. The 10-th, 25-th, and 50-th percentile results are in Appendix ~\ref{sec:app:res}. 

From the results in Table~\ref{tab:binding_overall}, it is easy to conclude:
1) \texttt{VD-Gen} significantly outperforms all other baselines in all metrics, with top-1 MM-PBSA Rank, demonstrating the superior performance of the proposed \texttt{VD-Gen}. 
2) MM-PBSA B.T shows that \texttt{VD-Gen} can generate more molecules with better MM-PBSA scores than the ground-truth ones, while baseline hardly can. 
3) In 3D Similarity results, \texttt{VD-Gen} also largely outperforms baselines, indicating that \texttt{VD-Gen} effectively learned the pocket-based 3D molecular generation and can generalize to unseen pockets. 
4) Although some baselines achieve good performance on Vina scores, like GraphBP and Pocket2Mol, their Vina* and MM-PBSA scores are very poor. We believe the re-docking in Vina fixes their generated 3D structures and then a good Vina score could be obtained. This result indicates that the previously widely used Vina score is not a good metric for pocket-based 3D molecular generation.

To summarize, the superior results on multiple evaluation metrics explicitly demonstrate the effectiveness of the proposed \texttt{VD-Gen}.


\prepar
\subsection{Ablation Study}

\prepar
\paragraph{Number of VPs}
VPs are used to approximate the distribution of molecule atoms. Intuitively, with more VPs, the approximation is more accurate. 
Therefore, we study how the number of VPs affects the final performance, the results are shown in Fig.~\ref{fig:ablation_vp_num}. From the result, it is clear that the number of VPs will affect the performance, and the results with more VPs are better. We also notice that the results are stable after 16 times of predicted molecular atoms. This indicates that it is not necessary to use too many VPs, we can use an appropriate number of VPs to achieve a trade-off between efficiency and performance.


\prepar
\paragraph{Number of movement iterations}
Iterative movement is critical in the \Name.
In Fig.~\ref{fig:ablation_equ_iteration} and  Fig.~\ref{fig:ablation_refine_iteration}, we benchmark the effectiveness of different iterations in \emph{Particle Movement} and \emph{Molecule Refinement}. For the results in Fig.~\ref{fig:ablation_equ_iteration}, we reduce the iterations $R_2$ to $0.25R_2$ in \emph{Molecule Refinement} stage, to better show the gain brought by \emph{Particle Movement} stage.
As shown in Fig.~\ref{fig:ablation_equ_iteration} and Fig.~\ref{fig:ablation_refine_iteration}, we can find more iteration iterations improve the final performance in both two stages.


\prepar
\paragraph{Effectiveness of Molecule Refinement}
The \emph{Molecule Refinement} stage is used to further refine the 3D molecule extracted by \emph{Molecule Extraction}. To examine how \emph{Molecule Refinement} affects the final performance, we benchmarked different iterations. As shown in Fig.~\ref{fig:ablation_refine_iteration}, we can find the results with more iterations are better. The result indicates the necessity of the \emph{Molecule Refinement} stage. \looseness=-1

\prepar
\paragraph{Effectiveness of Confidence Prediction} The pLDDT score is outputted at \emph{Confidence Prediction}, and used for selecting or ranking molecules, and we want to check its effectiveness. In particular, we calculate the correlation between 3D similarity and the pLDDT for the generated molecules on a pocket (PDBID 1LF2), and the result is shown in Fig.~\ref{fig:ablation_relevance_3d}. It is clear that with a larger pLDDT score, the corresponding 3D Similarity is better. This result indicates that the confidence score provided by \texttt{VD-Gen} is effective to select or rank the generated molecules. \looseness=-1

\prepar
\subsection{Case Study}

\begin{figure*}[t]
\centering
\includegraphics[width=0.95\linewidth]{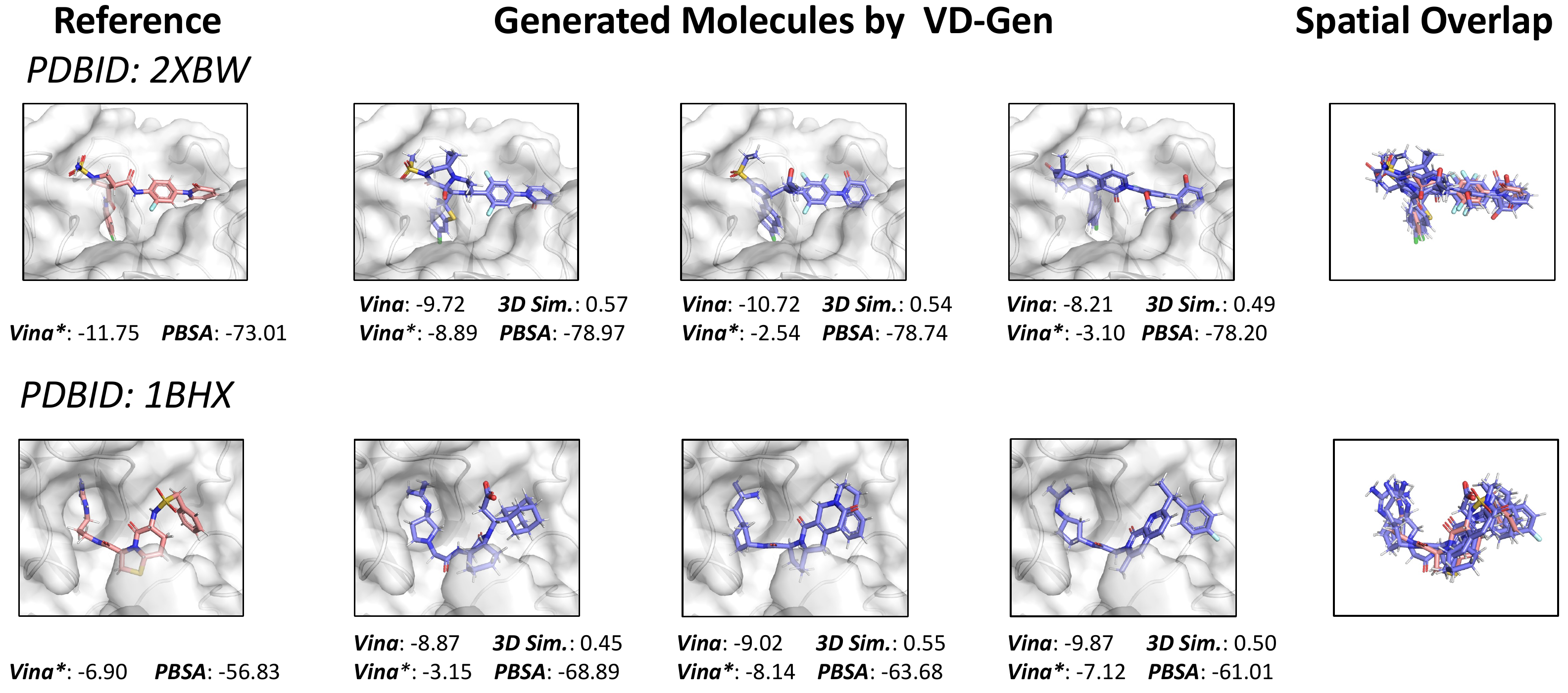}
\vspace{-5pt}
\caption{Generated molecules with high 3D similarity to the reference molecular and high PBSA scores for two protein pockets. 
Gray surfaces are the protein pockets. Green molecules are the ground truth molecules. Purple molecules are the molecules generated by \texttt{VD-Gen}.
Lower Vina score, lower PBSA score and higher 3D similarity indicate higher binding affinity. } \label{fig:show_case_gen}
\vspace{-8pt}
\end{figure*}%

Here, we selected two protein pockets from the test set to visualize the generated results of \texttt{VD-Gen} on pocket-based generation tasks. As shown in Fig~\ref{fig:show_case_gen}, for each pocket, 3 molecules (purple molecules in the middle column) with the top MM-PBSA scores are selected for display. These molecules are shown as they as, without any structural post-processing. Green molecules are the ground truth molecules, and the rightmost column is the spatial overlapping of the generated molecules and the original molecule. \looseness=-1

In the first case (PDBID: 2XBW), the protein pocket has a pit deep inside the protein (bottom left of the image), the volume of which can accommodate about one benzene ring. It is a challenging task due to the small size of the pit and the long distance from the center of the whole pocket. We can see that the molecules generated by \texttt{VD-Gen} have successfully grown fragments within the pit. On the other hand, the three generated molecules have good 3D similarity with the original molecules, and the MM-PBSA score is good, the Vina scores of the original molecule are much better than those of the three generated molecules. If we only use Vina to pick molecules, It may lead to not picking good molecules. \looseness=-1

In the second case (PDBID: 1BHX), the protein pocket is bulky, which requires the generation of protein-interacting fragments at both ends of the protein pocket, and connecting the two ends together by a molecular backbone, we can see the original molecule is long and distorted, making it a challenging prediction task. We see that the molecules generated by \texttt{VD-Gen} replicate the shape of the original molecules well, filling the uneven protein pockets well. All three molecules have good 3D similarity and MM-PBSA scores. \looseness=-1


From these cases in Fig~\ref{fig:show_case_gen}, we can see that \texttt{VD-Gen} has demonstrated good generation capabilities on different types of challenging molecular generation tasks.
For example, the generated molecules can fill deep pockets, follow the trend of large pockets, or match the special structure of the pockets, and the 3D similarity between the generated molecule and the molecule in the original crystal structure is high. 
On the other hand, we can see that the MM-PBSA score and 3D similarity maintain good consistency in evaluating the quality of generated molecules, while the Vina score fails in some cases, which indicates that it is unreasonable to select molecules based on the Vina score alone. \looseness=-1

\prepar
\subsection{Visualization} 

\begin{figure*}[ht]
\centering
\includegraphics[width=0.94\linewidth]{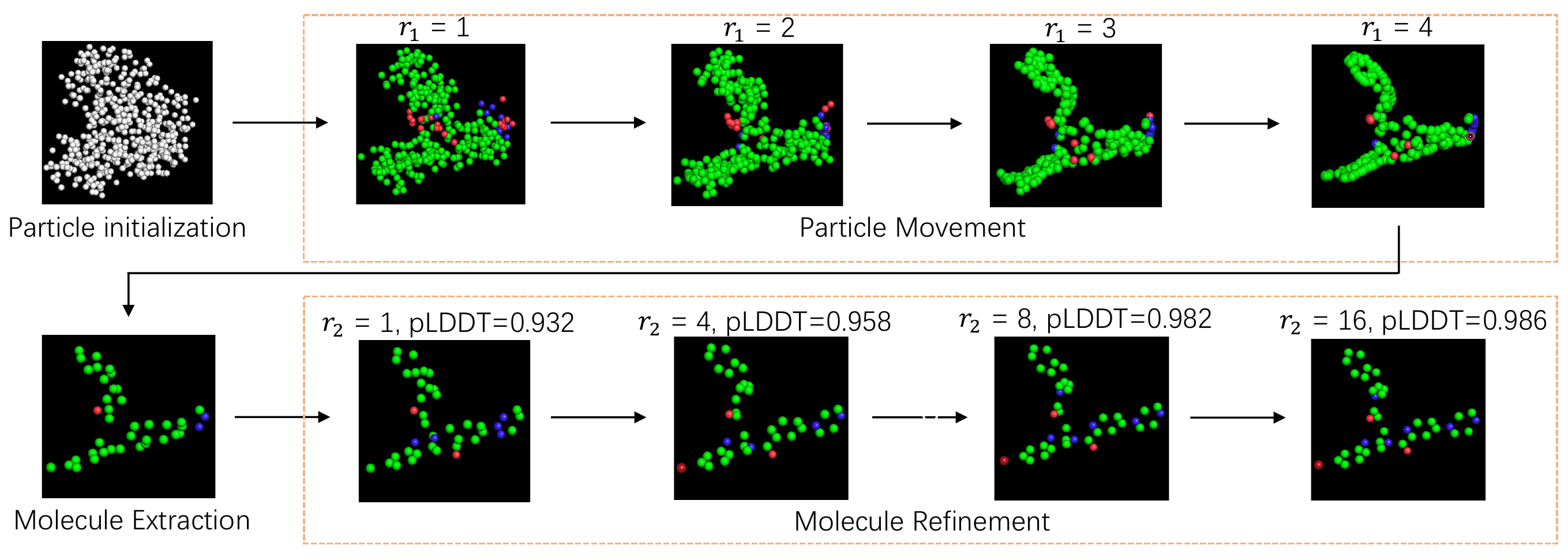}
\vspace{-8pt}
\caption{An example to demonstrate the output results of \texttt{VD-Gen}'s different stages, $r_1$ ($r_2$) is the iteration number of \emph{Particle Movement} (\emph{Molecule Refinement}). The increasing pLDDT scores in the pipeline indicate the effectiveness of \Name. } \label{fig:move_vase}
\vspace{-8pt}
\end{figure*}%

To better understand how \Name generates the 3D molecules, we also provide a visualization for the stages in \Name, shown in Fig~\ref{fig:move_vase}. At the beginning, the VPs are uniformly initialized inside the protein pocket. Then in \textit{Particle Movement} stage, with more iterations, VPs gradually aggregate into several clusters, to approach the positions of molecular atoms. Then in \textit{Molecule Extraction} stage, a 3D molecule with fewer atoms is extracted. Then in \textit{Molecule Refinement} stage, the extracted 3D molecule is further refined,  toward 3D positions with better pLDDT scores. \looseness=-1

\vspace{-5pt}
\section{Related Work}
\vspace{-5pt}

\paragraph{Ligand-Based Molecular Generation}
Early works focused on ligand-based molecular generation, took a set of molecules as training data, and generated molecules based on the learned distribution of training data. And these methods mainly represented molecules as 1D SMILES strings and 2D molecular graphs, and used VAEs~\cite{DBLP:conf/icml/KusnerPH17, DBLP:conf/iclr/DaiTDSS18, winter2019efficient, griffiths2020constrained, dollar2021attention, DBLP:journals/jcisd/OliveiraSQ22}, GANs~\cite{guimaraes_objective-reinforced_2018,sanchez-lengeling_optimizing_nodate}, flow models~\cite{shi_graphaf_2020} for one-shot generation, RNNs~\cite{DBLP:journals/jcheminf/OlivecronaBEC17, DBLP:journals/corr/BjerrumT17, segler2018generating, DBLP:journals/corr/abs-2112-03041}, reinforcement learning approaches\cite{you_graph_2019, jin_multi-objective_2020} for step-by-step generation. And some works~\cite{li_deepscaffold_2020, lim_scaffold-based_2020, chemicalVAE} tried to preserve structural features like molecular scaffolds, or physicochemical properties like QED, to gain better generated molecules compared to randomly generation. However, those methods did not take the binding affinity against a specific protein pocket as a target directly thus the generated molecules hardly worked well in real-world tasks. Some recent works~\cite{nesterov_3dmolnet_2020, simm_reinforcement_2020,hoogeboom2022equivariant,wu2022diffusion} also tried the ligand-based 3D molecular generation. 

\vspace{-6pt}
\paragraph{Pocket-Based Molecular Generation} Due to the importance of binding affinity in drug design, recent works involved the information of protein pockets for molecular generation. Early attempts~\cite{skalic_target_2019, xu_novo_2021} encoded pocket information and took it as a condition to generate molecules in SMILES strings or molecular graphs. However, since the binding affinity depends on the spatial positions of pocket and molecule, the latter works paid more effort in generating molecules with 3D spatial structures. Some works~\cite{ragoza_generating_2022}, recognized as molecular 3D density grid generation, converted pockets and molecules into 3D density grids, and applied 3D convolutional models like processing images. But as the pocket cavity is large, the positions of pockets and molecules are coarse-grained in 3D density grids and it leads to information loss and hard to generate fine-grained molecules. Besides, it is not end-to-end since the conversion from 3D density to 3D coordinates is required and usually causes additional accuracy loss. 
Some other works~\cite{sbdd, liu_generating_2022, peng_pocket2mol_2022}, recognized as auto-regressive 3D molecular generation, sampled/generated atoms in 3D space one by one to form a molecule. Suffering from the large space of continuous 3D positions, it is quite inefficient. 
Besides, unlike the sequential nature in text, the atoms in a molecule do not have a sequential order. That is, we do not know which atoms should be generated first, and thus, using auto-regressive generation for 3D molecules is not reasonable.

\prepar
\section{Conclusion}
\vspace{-5pt}

In this paper, we propose \texttt{VD-Gen}, a novel pocket-based 3D molecular generation pipeline, to generate fine-grained 3D molecules with good binding affinities against the protein pocket end-to-end. In particular, many virtual particles are first randomly distributed in the pocket cavity, and then are iteratively moved to approximate the distribution of molecular atoms from the training data. Then, a 3D molecule could be extracted by deep models from these virtual particles. Next, the atoms in the extracted molecule are continually refined by iterative movement again, and a high-quality 3D molecule with fine-grained coordinates could be obtained. Finally, a confidence score will be calculated for the generated molecule for the need of selecting or ranking. 
Experiment results demonstrate that \texttt{VD-Gen} can generate molecules with higher binding affinities to protein pockets and more accurate 3D binding structures than other baselines. Ablation study, case study and visualizations are further provided to demonstrate the effectiveness of \texttt{VD-Gen}.\looseness=-1


\bibliography{reference}
\bibliographystyle{unsrt}
\newpage

\appendix

\onecolumn

\section{VD-Gen details}
\label{Appendix:VD}

\begin{table}[H]
  \centering
  \small
  \vspace{-20pt}
  \caption{Symbols used in this paper.} \label{app:table:symbol}
  \vskip 0.15in
    \begin{tabular}{l|l}
    \toprule
    Symbol & Meaning\\
    \hline
    $\mathbf{P}$ & the set of atoms in the pocket \\
    $\mathbf{V}_{r}$ & the set of virtual particles (VPs) that are generated at the $r$-th iteration in \textit{Particle Movement}\\
    $\mathbf{W}_{r}$ & the set of virtual particles (VPs) that are generated at the $r$-th iteration in \textit{Molecule Refinement}\\
    $\mathbf{C}_p$ & the gridded 3D cubic of pocket atoms \\
    $\mathbf{C}_m$ & the predicted cubic gridded cubic \\
    $\mathbf{C}^g_m$ & the groud truth label for the gridded cubic \\
    $\mathbf{G}$ & the set of ground-truth atoms\\
    $\bm{x}_i^p$ & the $i$-th pocket atom's type (one-hot)\\
    $\bm{y}_i^p$ & the $i$-th pocket atom's coordinate\\
    $\bm{x}_i^g$ & the $i$-th ground-truth atom's type (one-hot)\\
    $\bm{y}_i^g$ & the $i$-th ground-truth atom's coordinate\\
    $\bm{x}_i^r$ & the $i$-th VP's type (one-hot) at the $r$-th iteration in \textit{Particle Movement}\\
    $\bm{y}_i^r$ & the $i$-th VP's coordinate at the $r$-th iteration in \textit{Particle Movement}\\
    $\bar{\bm{x}}_i^r$ & predicted atom type distribution of $i$-th VP at the $r$-th iteration\\
    $a_i$ & The index of assigned target atom for the $i$-th VP in \textit{Particle Movement}\\
    $b_i$ & The index of assigned target atom for the $i$-th VP in \textit{Molecule Refinement}\\
    ${\bm{d}}_{ij}^r$ & the predicted distance of the $i$-th and $j$-th VP pair at the $r$-th iteration \\
    $\bm{d}_{a_i,a_j}^g$ & the ground-truth distance of the $i$-th and $j$-th VP pair \\
    ${\bm{c}}_{ij}^r$ & the predicted (ground-truth) distance of the $i$-th VP and the $j$-th pocket atom at the $r$-th iteration\\
    $\bm{c}_{a_i,j}^g$ & the ground-truth 
 distance of the $i$-th VP and the $j$-th pocket atom.\\
 $\hat{\bm{x}}_i^r$ & the $i$-th VP's type (one-hot) at the $r$-th iteration in \textit{Molecule Refinement}\\
    $\hat{\bm{y}}_i^r$ & the $i$-th VP's coordinate at the $r$-th iteration in in \textit{Molecule Refinement}\\
    $\bm{q}$ & the pair representation of VP pair \\
    $\bm{q}^l$ & the pair representation of VP pair at $l$-th layer \\
    $\bm{q}^P$ & the pair representation of pocket atom pair \\
    $\bm{q}^C$ & the pair representation of VP and pocket pair \\
    $\bar{\bm{s}}$ & the predicted distance between VP and its target which is used to filter VP.\\
    $\bar{\bm{r}}$ & predicted probability of merging type of VP pair\\
    $\bar{m}$ & the predicted atom number \\
    $k_{vp}$ & times of the number of atom, uses in \textit{Particle Initialization} \\
    $\bm{w}_i$ & The indices of VPs in the $i$-th cluster in \emph{Molecule Extraction} \\
    $\bm{h}$ & the node representation of VP \\
    $\bm{h}^l$ & the node representation of VP at $l$-th layer \\
    $\bm{h}^V$ & the node representation of VP in \textit{Particle Movement} \\
    $\bm{q}^V$ & the pair representation of VP pair in \textit{Particle Movement} \\
    $\bm{h}^W$ & the node representation of VP in \textit{Molecule Refinement} \\
    $\bm{q}^W$ & the pair representation of VP pair in \textit{Molecule Refinement} \\
    $\bm{h}^P$ & the node representation of pocket atom \\
    $h_{an}$ & the model to predict atom number\\
    $\bm{\theta}_{an}$ & the model parameter to predict atom number\\
    $h_{pi}$ & the model to predict pocket cavity in \textit{Particle Initialization}\\
    $\bm{\theta}_{pi}$ & the model to predict pocket cavity in \textit{Particle Initialization}\\
    $\bm{\theta}_{pm}$ & the model parameter in \textit{Particle Movement}\\
    $\bm{\theta}_{me}$ & the model parameter in \textit{Molecule Extraction}\\
    $\bm{\theta}_{mr}$ & the model parameter in \textit{Molecule Refinement}\\
    $\bm{\theta}_{cp}$ & the model parameter in \textit{Confidence Prediction}\\
    $f_{pm}$ & the SE(3) backbone model, return types and coordinates of atoms/particles in \textit{Particle Movement} \\
    $f_{me}$ & the SE(3) backbone model, return types and coordinates of atoms/particles in \textit{Molecule Refinement} \\
    $L$ & the number of layers \\
    \bottomrule
    \end{tabular}
\vskip -0.1in
\end{table}

\subsection{Details of the SE(3) backbone model} \label{app:backbone}

In Fig~\ref{fig:model_stru} we show the structure of the SE(3) backbone model used in \Name. "Repr.", "Attn." and "Dist." are the abbreviations of "Representation", "Attention" and "Distance", respectively. On the left is the pocket encoder, which first uses an atom-type embedding to encode the pocket atom type and a Gaussian kernel to encode the pair-wise distances between pocket atom pairs. In each layer of the pocket encoder, a self-attention layer is used. On the right is the encoder for VPs, which also uses an atom-type embedding and a Gaussian kernel to encode the particle type and the pair-wise distances between VPs. To interact with the pocket encoder, another Gaussian kernel is used to encode the pair-wise distances between VPs and pocket atoms. In each layer of the VP encoder, before the self-attention layer, a particle-pocket attention layer is used to interact with the pocket encoder.

We describe the components in the backbone model in the following paragraphs. Besides, we also describe the overall pipeline of the backbone model in the Alg.~\ref{alg:backbone}. For simplicity, layer normalization is not shown in the equations and algorithms.

\paragraph{Gaussian kernel} The pair-type aware Gaussian kernel~\cite{shuaibi2021rotation, zhou2022uni} is denoted as: 
\begin{equation}
\boldsymbol{p}_{ij} = \{ \mathcal{G}(\mathcal{A}(d_{ij}, t_{ij} ; \boldsymbol{a}, \boldsymbol{b}), \mu^k, \sigma^k) | k \in [1, D]\}, 
\mathcal{A}(d, r; \boldsymbol{a}, \boldsymbol{b}) = a_r d + b_r,
\end{equation}
where $\mathcal{G}(d, \mu, \sigma)= \frac{1}{\sigma\sqrt{2\pi}} e^{-\frac{(d-\mu)^2}{2\sigma^2}}$ is a Gaussian density function with parameters $\mu$ and $\sigma$, $d_{ij}$ is the Euclidean distance of atom pair $ij$, and $t_{ij}$ is the pair-type of atom pair $ij$. 
$\mathcal{A}(d_{ij}, t_{ij}; \boldsymbol{a}, \boldsymbol{b})$ is the affine transformation with parameters $\boldsymbol{a}$ and $\boldsymbol{b}$, it affines $d_{ij}$ corresponding to its pair-type $t_{ij}$. 

\paragraph{Pair representation} 
Pair representation \cite{zhou2022uni} is used to further enhance the 3D spatial encoding. The update of pair representation is via the multi-head Query-Key product results in self-attention. 
\begin{equation}
\label{eq:pair_update}
\boldsymbol{q}^{l+1}_{ij} = \boldsymbol{q}^{l}_{ij} + \{\frac{\bm{h}_i^l \boldsymbol{W}^Q_{l,h}(\bm{h}_j^l \boldsymbol{W}^K_{l,h})^T}{\sqrt{d}} | h \in [1, H] \},
\end{equation}
where $\bm{h}_i^l$ is the atom/node representation of the $i$-th atom at $l$-th layer, $\boldsymbol{q}^{l}_{ij}$ is the pair representation of atom pair $ij$ in $l$-th layer, $H$ is the number of attention heads, $d$ is the dimension of hidden representations, and $\boldsymbol{W}^Q_{l,h}$ ($\boldsymbol{W}^K_{l,h}$) is the projection for Query (Key) of the $l$-th layer $h$-th head.

To leverage 3D information in the atom representation, pair representation is used in self-attention.
\begin{equation}
\label{eq:pair2atom}
\begin{aligned}
\bm{h}_{i}^{l+1,h} &= \text{softmax}(\frac{\bm{h}_i^l \boldsymbol{W}^Q_{l,h}(\bm{h}_j^l\boldsymbol{W}^K_{l,h})^T}{\sqrt{d}} + \boldsymbol{q}_{ij}^{l,h})\bm{h}_j^l \boldsymbol{W}^V_{l,h}, \\
\bm{h}_i^{l+1} &= \text{concat}_{h}(\bm{h}^{l+1,h}_{i}),
\end{aligned}
\end{equation}
where $\boldsymbol{W}^V_{l,h}$ is the projection of Value of the $l$-th layer $h$-th head.  

\paragraph{Particle-Pocket Attention}

The Particle-Pocket Attention can be denoted as the following:
\begin{equation}
\begin{aligned}
\label{eq:gated}
\bm{h}_{i}^{l+1,h} &= \text{softmax}\left(\frac{\bm{h}_i^l \boldsymbol{W}^{P,Q}_{l,h}(\bm{h}_j^P\boldsymbol{W}^{P,K}_{l,h})^T}{\sqrt{d}} + \boldsymbol{q}_{ij}^{C,l,h}\right)\bm{h}_j^P \boldsymbol{W}^{P,V}_{l,h}, \\
\bm{h}_i^{l+1} &= \text{concat}_{h}(\bm{h}^{l+1,h}_{i}), \\
\bm{h}_i^{l+1} &= \bm{h}_i^{l} + g_1 \cdot \bm{h}_i^{l+1} +  g_2 \cdot \text{MLP}(\bm{h}_i^{l+1}),
\end{aligned}
\end{equation}
where $g_1$ and $g_2$ are learned parameters with initialized value 0, $\bm{h}_j^P$ is the representation of the $j$-th pocket atom, $\boldsymbol{q}^{C,l,h}_{ij}$ is the pair representation of particle-pocket pair $ij$ in $l$-th layer $h$-th head,
MLP is a full-connected network with one hidden layer. $\boldsymbol{W}^{P,Q}_{l,h}$, $\boldsymbol{W}^{P,K}_{l,h}$, and $\boldsymbol{W}^{P,V}_{l,h}$ are learnable projections for Query, Key and Value.

\paragraph{SE(3)-equivariance coordinate} Following \cite{zhou2022uni}, the head could be denoted as:

\begin{equation}
\begin{aligned}
    \boldsymbol{y}_i^{r+1} &= \boldsymbol{y}_i^r + \sum_{j=1}^{n}\frac{(\boldsymbol{y}_i^r - \boldsymbol{y}_j^r)z_{ij}}{n},
    z_{ij} &= \text{ReLU}((\boldsymbol{q}^L_{ij} - \boldsymbol{q}^0_{ij})\boldsymbol{U}_1)\boldsymbol{U}_2,\\
\end{aligned}
\end{equation}
where $n$ is the number of total atoms, $L$ is the number of layers in model, $\boldsymbol{y}_i^r \in \mathbb{R}^3$ is the input coordinate of $i$-th atom, and $\boldsymbol{y}_i^{r+1} \in \mathbb{R}^3$ is the output coordinate of $i$-th atom,  $\boldsymbol{U}_1 \in \mathbb{R}^{H \times H}$ and $\boldsymbol{U}_2 \in \mathbb{R}^{H \times 1}$ are the projection matrices to convert pair representation to scalar. Note that we also use the predicted coordinates to calculate the distance between VPs as the predicted distance $\bm{d}^r_{ij}$ in Equation~\ref{equ:atom_coord_loss}.

\paragraph{Atom Type Prediction Head}
We use a non-linear head with two layers to predict the atom type based on the atom representation  in the last layer of the particle encoder:
\begin{equation}
\bar{\bm{x}}_i = \text{MLP}(\bm{h}^L_i)
\end{equation}
where $\bm{h}^L_i$ is the atom representation, $L$ is the number of layers of the particle encoder,

\begin{figure}[H]
\vspace{-10pt}
\begin{algorithm}[H]
\caption{Backbone\_Update}\label{alg:backbone}
\begin{algorithmic}[1]
\small
\Require {$\mathbf{P}$: pocket atoms, $\mathbf{V}_r$: virtual particles}
\State $\bm{h}^{P, 0} \gets \text{Atom\_Type\_Embedding}(\mathbf{P})$ \Comment{Embeddings from atom types}
\State $\bm{q}^{P, 0} \gets \text{Gaussian\_Kernel}(\text{Dist\_Matrix}(\mathbf{P}, \mathbf{P}))$ \Comment{Get invariant spatial positional embedding}
\State  \textit{{\color{blue}\# Update Pocket Encoder}}
\For{$l \in [1,...,L)$} 
\State $\bm{h}^{P, l}, \bm{q}^{P,l}   \gets \text{Self\_Attn}(\bm{h}^{P, l-1}, \bm{q}^{P,l-1}) $) \Comment{Update by self attention}
\State $\bm{h}^{P, l} \gets \text{MLP}(\bm{h}^{P, l})$ \Comment{Update by Feed-Forward-Network}
\EndFor

\State $\bm{h}^P \gets \bm{h}^{P, L}$

\State $\bm{h}^{0} \gets \text{Atom\_Type\_Embedding}(\mathbf{V}_r)$ \Comment{Embeddings from atom types}
\State $\bm{q}^{0} \gets \text{Gaussian\_Kernel}(\text{Dist\_Matrix}(\mathbf{V}_r, \mathbf{V}_r))$ \Comment{Get invariant spatial positional embedding}
\State $\bm{q}^{C,0} \gets \text{Gaussian\_Kernel}(\text{Dist\_Matrix}(\mathbf{V}_r, \mathbf{P}))$ \Comment{Get invariant spatial positional embedding of particle-pocket pairs}
\State  \textit{{\color{blue}\# Update Particle Encoder}}
\For{$l \in [1,...,L)$} 
\State $\bm{h}^{l}, \bm{q}^{l}   \gets \text{Self\_Attn}(\bm{h}^{l-1}, \bm{q}^{l-1}) $) \Comment{Update by self attention}
\State $\bm{h}^{l} \gets \text{MLP}(\bm{h}^{l})$ \Comment{Update by Feed-Forward-Network}
\State  \textit{{\color{blue}\# Only enabled at every 4-layer}}
\If {$l \mod 4$ == $0$}
\State $\bm{h}^{l}, \bm{q}^{C, l}   \gets \text{Particle\_Pocket\_Attn}(\bm{h}^{l}, \bm{h}^P, \bm{q}^{C, l-1}) $) \Comment{Update by Particle-Pocket Attention}
\EndIf
\EndFor

\State $\bar{\bm{x}}^{r+1} \gets \text{Atom\_Type\_Head}(\bm{h}^L)$ \Comment{Atom Type Prediction}
\State $\bm{x}^{r+1} \gets \text{sample}(\bar{\bm{x}}^{r+1})$ \Comment{Sample an atom type based on predicted probability}
\State $\bm{y}^{r+1} \gets \text{SE(3)\_Head}(\bm{y}^{r}, \bm{q}^{L})$ \Comment{Coordinate update} \\

\Return  $\mathbf{V}_{r+1} = \{\bm{x}^{r+1}, \bm{y}^{r+1}\}$, $\bm{h}^L$, $\bm{q}^L$

\end{algorithmic}
\end{algorithm}
\vspace{-15pt}
\end{figure}

\begin{figure}[t]
\centering
\includegraphics[width=0.8\linewidth]{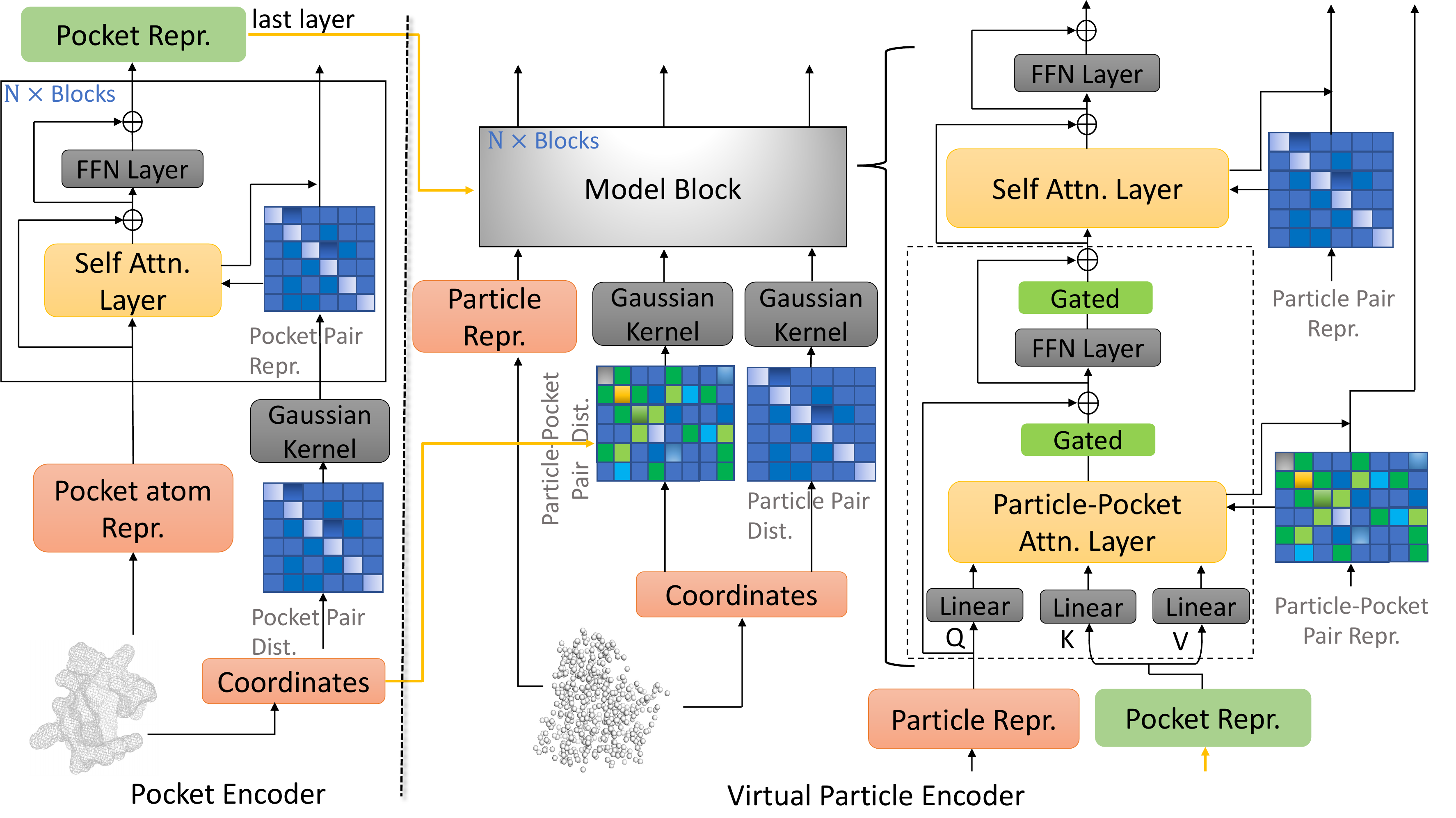}%
\caption{The backbone model used in \texttt{VD-Gen}. Details are in Appendix~\ref{app:backbone} and Alg.~\ref{alg:backbone}.} \label{fig:model_stru}
\end{figure}%

\subsection{Training Loss in VD-Gen}
\label{Appendix:loss}

\paragraph{Binning regression}
As described in Sec.~\ref{sec:train}, we convert several regression tasks into classification tasks by binning. Then, the training loss of classification can be written as:
\begin{equation}
\small
\mathcal{L}_{binning\_regression} = \frac{1}{n} \sum_{i=1}^n\text{NLL}( \bar{\bm{z}}_i, \bm{z}_i) 
\end{equation}
where $\bar{\bm{z}}_i$ is the predicted probability distribution of bins, and $\bm{z}_i$ represents the one-hot vector of the target bin. 

During inference, the predicted value can be calculated from the predicted distribution over bins. 
\begin{align}
\bar{z}_i = \sum^{n_{bin}}_{k=1} (bin\_val_{k}) \bar{\bm{z}}_i[k],
\end{align}
where $n$ is the number of samples, $bin\_val_k$ is the bin value of the $k$-th bin, $n_{\text{bin}}$ is the number of bins, $l$ is the size of each bin and $\bar{\bm{z}}_i[k]$ is the predicted probability of the $k$-th bin. Notably, the bin value is not the bin boundary value, it is the average of left and right boundaries.

\paragraph{Focal loss}
There are also several Focal losses used in Sec.~\ref{sec:train}. Formally, it can be denoted as:
\begin{equation}
\small
\mathcal{L}_{Focal} =  \frac{1}{n} \sum^n_{i=1}\sum^K_{k=1} -\bm{z}^k_{i}\log \bar{\bm{z}}^k_{i}{\color{blue}(1-\bar{\bm{z}}^k_{i})^{\gamma}},
\end{equation}
where $n$ is the number of samples, $K$ is the number of types, $\bm{z}_{i}$ is the one-hot vector of target type, $\bar{\bm{z}}_{i}$ is the predicted probability of type, the {\color{blue} blue} part is from focal loss~\cite{lin2017focal}, and $\gamma$ are hyper-parameters to balance classes. In this paper $\gamma$ is set to $2$.

\paragraph{LDDT}

LDDT score is widely used in protein structure prediction \cite{mariani2013lddt}, and it can be denoted as:
\begin{equation}
\begin{aligned}
\text{LDDT} &=\frac{1}{n}\sum_{i=1}^n \sum_{j\neq i}  \frac{1}{4} ((\text{err}_{ij} < 0.5) + (\text{err}_{ij} < 1.0) \\ & + (\text{err}_{ij} < 2.0) + (\text{err}_{ij}) < 4.0),\\ 
\end{aligned}
\end{equation}
\begin{align}
\text{where } \text{err}_{ij} &= \text{L1}(||\hat{\bm{y}}^{R_2}_i - \hat{\bm{y}}^{R_2}_j||_2, ||\hat{\bm{y}}^g_i - \hat{\bm{y}}^g_j||_2),
\end{align}
$\hat{\bm{y}}^{R_2}_i$ is the predicted coordinate of $i$-th particle after  \textit{Molecule Refinement}, and $\hat{\bm{y}}^g_i$ is its ground truth coordinate.

\subsection{VD-Gen Overall Algorithm}

We also summarize the overall inference pipeline of \texttt{VD-Gen} in the Alg. \ref{alg:overall_infer}. First the pocket encoder and a head are used to predict atom number based on the pocket representation. Then a 3D U-net model is used to predict pocket cavity based on the pocket atoms. Then the main algorithm mainly relies on the function "Iter\_Move", which iteratively moves the VPs. Both \emph{Particle Movement} and \emph{Molecule Refinement} use "Iter\_Move". In \emph{Molecule Extraction}, several heads are used to predict the filtered probability and the pair merging probability based on the particle representation $\bm{h}^V$ and the pair  representation between particles $\bm{q}^V$. Based on these predictions, "Molecule\_Extraction" is used to extract the merged VPs as in Algorithm~\ref{alg:merge}. 

The training pipeline is very similar, except for the following differences:
\begin{itemize}
    \item For efficiency purposes, $R_1$ and $R_2$ are all sampled from $1 \sim 4$, and the gradient backward is only enabled in the last iteration.
    \item For efficiency purposes, in "Molecule\_Extraction", teacher-forcing merging (without binary search) is used during training. 
    This is, rather than predicting pair-wise merging probabilities and the atom number, we directly used their ground truth values. 
    \item The loss functions are enabled to get gradients for training.
\end{itemize}

\begin{figure}[ht]
\vspace{-10pt}
\begin{algorithm}[H]
\caption{VD-Gen Inference Pipeline}\label{alg:overall_infer}
\begin{algorithmic}[1]
\small

\Require {$R_1$,  $R_2$: iterations in \textit{Particle Movement} and \textit{Molecule Refinement}, $k_{vp}$: times of the number of atom, $\mathbf{P}$: pocket atoms with types and positions, $\mathbf{C}_p$: the gridded 3D cubic of pocket atoms,
$h_{an}$: the model to predict atom number, 
$\bm{\theta}_{an}$: the model parameter to predict atom number,
$h_{pi}$: the model to predict pocket cavity in \textit{Particle Initialization}, 
$\bm{\theta}_{pi}$: the model parameter to predict pocket cavity in \textit{Particle Initialization}, 
$\bm{\theta}_{pm}$, $\bm{\theta}_{me}$, $\bm{\theta}_{mr}$, $\bm{\theta}_{cp}$,: model parameters in \textit{Particle Movement}, \textit{Molecular Extraction}, \textit{Molecular Refinement} and \textit{Confidence Prediction}  }\\

\State  \textit{{\color{blue}\# Iterative Movement}}
\Function{Iter\_Move}{$R$, $\mathbf{V}_{0}$, $\mathbf{P}$, $\bm{\theta}$}:
\For{$k \in [1,...,R]$}
\State $\mathbf{V}_k, \bm{h}^L, \bm{q}^L, \gets \text{Backbone\_Update}(\mathbf{V}_{k-1}, \mathbf{P}; \bm{\theta} $) \Comment{backbone model update as in Alg.~\ref{alg:backbone}}
\EndFor
\State \textbf{output} $\mathbf{V}_{R}, \bm{h}^L, \bm{q}^L$ 
\EndFunction

\State \textit{{\color{blue}\# Particle Initialization}}
\State $\bar{m} = {h}_{an}(\mathbf{P}, \bm{\theta}_{an})$ \Comment{Predict atom number}
\State $ \mathbf{C}_m = h_{pi}(\mathbf{C}_p, \bm{\theta}_{pi})$ \Comment{use 3D U-net model to predict the gridded cubic}
\For{$i$ in [1,...,$\bar{m}k_{vp}$]}
\State $\bm{x}^0_i \gets \text{one\_hot([MASK])}$ \Comment{The types of VPs are initialized as a meaningless [MASK] type}
\State $\bm{y}^0_i \gets \text{Uniform}( \mathbf{C}_m[\mathbf{C}_m == 1])$ \Comment{the grids in the predicted gridded cubic with voxel value $1$ are taken as cavity and the initial coordinates of VPs are uniformly sampled from the cavity space}
\EndFor
\State $\mathbf{V}_0 = \{(\bm{x}^0_i, \bm{y}^0_i)\}^n_{i=1}$\\

\State  \textit{{\color{blue}\# Particle Movement }}
\State $\mathbf{V}_{R_1}, \bm{h}^V, \bm{q}^V \gets \text{Iter\_Move}( R_1, \mathbf{V}_{0}, \mathbf{P}, \bm{\theta}_{pm})$ \Comment{Predict coordinates and types with iterative movement}\\

\State \textit{{\color{blue}\# Molecule Extraction}}
\State $\bm{\bar{s}} \gets \text{Predict\_Distance}(\bm{h}^V; \bm{\theta}_{me})$ \Comment{Predict the distance between VPs and the targets to filter particles}
\State $ \bar{\bm{r}} \gets \text{Predict\_Merge}(\bm{q}^V; \bm{\theta}_{me})$ \Comment{Predict merging matrix}

\State $\mathbf{W}_{0} \gets \text{Molecule\_Extraction}(\mathbf{V}_{R_1}, \bar{m}, \bar{\bm{r}}, \bar{\bm{s}})$\\ \Comment{Filter and Merge VPs as in Alg.~\ref{alg:merge}  using predicted merging matrix, atom number and predicted distance}\\

\State \textit{{\color{blue}\# Molecule Refinement}}
\State $\mathbf{W}_{R_2}, \bm{h}^W, \bm{q}^W, \gets \text{Iter\_Move}(R_2,\mathbf{W}_{0}, \mathbf{P}; \bm{\theta}_{mr})$ \Comment{Refine the coordinates and types}

\State \textit{{\color{blue}\# Confidence Prediction}}
\State Pred\_LDDT $\gets$ Predict\_Confidence($\bm{h}^W$; $\bm{\theta}_{cp}$) \Comment{Predict LDDT}\\
\Return  $\mathbf{W}_{R_2}$, Pred\_LDDT \Comment{Return the final positions and types, and the confidence score }
\end{algorithmic}
\end{algorithm}
\vspace{-10pt}
\end{figure}

\subsection{Molecule Extraction algorithm}
The detail of merging VPs into atoms are shown in Alg~\ref{alg:merge}. In particular, a binary search is used to find a merging threshold. During training, teacher-forcing merging is used for reducing the training cost (without binary search). This is, rather than predicting pair-wise merging probabilities and the atom number, we directly used their ground truth values. 
During inference, the binary search is used. Besides, considering the error in atom number prediction, we try a range ($\pm10$) of atom numbers, and select from them based on their confidence scores.

\begin{figure}[H]
\begin{algorithm}[H]
\caption{Molecule Extraction Algorithm}\label{alg:merge}
\begin{algorithmic}[1]
\small
\Require {$\mathbf{V}_{R_1} = \{(\bm{x}_i^{R_1}, \bm{y}_i^{R_1})  \}^n_{i=1}$: virtual particles at $R_1$ iterations, $\bar{m}$: the predicted atom num,  $\bar{\bm{r}} = \{(\bar{\bm{r}}_{ij})\}_{i=1,j=1}^{n \times n}$: the predicted merging probability matrix, $\bar{\bm{s}} = \{\bar{{s}}_i \}_{i=1}^n$: the predicted distance between VPs and target positions, $\zeta$: the filtering threshold}
\State  \textit{{\color{blue}\# Set the merge type between particles with the particle to be filtered to $0$}}
\For {$i$ in $[1,...,n]$ } 
\State  \textit{{\color{blue}\# Filter the VPs according to predicted distance }}
\If{ $\bar{s}_i > \zeta$} 
\State $\bar{\bm{r}}[:,i] \gets 0$ \Comment{Set $\{\bar{\bm{r}}_{i,j}\}^n_{j=1}$ to $0$}
\State $\bar{\bm{r}}[i, :] \gets 0$ \Comment{Set $\{\bar{\bm{r}}_{j,i}\}^n_{j=1}$ to $0$}
\EndIf
\EndFor

\State $\text{high} \gets \text{max}(\{(\bar{\bm{r}}_{ij})\}_{i=1,j=1}^{n \times n})$
\State $\text{low} \gets \text{min}(\{(\bar{\bm{r}}_{ij})\}_{i=1,j=1}^{n \times n})$
\State $\text{mid} \gets \frac{\text{low} + \text{high}}{2}$
\State  \textit{{\color{blue}\# using the binary search to find the  threshold }}
\While{$\text{low} < \text{high}$} 
\State $\bm{r}_{ij} \gets \bar{\bm{r}}_{ij} > \text{mid}$
\State $\mathbf{W}_0 \gets [], \bm{w} \gets [], m \gets 0$
\State  \textit{{\color{blue}\# Greedy merge based a random order }}
\For {$i$ in \textbf{random\_perm}$(1, n)$ } 
\State $\bm{w}_i \gets [],$
\For {$j$ in $[1,...,n]$ }
\If{$\bm{r}_{ij} = \text{True}$} 
\State $\bm{r}[:,j] \gets \text{False}$ \Comment{the merged particle will not be merged again}
\State $\bm{w}_i.\text{add}(j)$ \Comment{add the particle indices into the $i$-th cluster}
\EndIf
\EndFor

\If{ $\text{len}(\bm{w}_i) > 0$}
\State $\hat{\bm{x}}_m^0 \gets \text{Uniform} (\{\bm{x}^{R_1}_k | k \in \bm{w}_i \})$ \Comment{Sample atom type from the merging list}
\State $\hat{\bm{y}}_m^0 \gets \text{Mean} (\{\bm{y}^{R_1}_k | k \in \bm{w}_i \})$ \Comment{the weighted average position according to the predicted distance is atom position after merging }
\State $\mathbf{W}_0.\text{add}((\hat{{\bm{x}}}^0_m,  \hat{{\bm{y}}}^0_m))$ \Comment{add the atom to the set }
\State $m \gets m + 1$ \Comment{Count the number of clusters}
\EndIf

\EndFor
\State  \textit{{\color{blue}\# find the threshold }}
\If{$m = \bar{m}$} 
\State break
\Else
\If{$m < \bar{m}$} 
\State $\text{low} \gets \text{mid}$ \Comment{
too many particles are merged, the threshold needs to be increased}
\Else
\State $\text{high} \gets \text{mid}$ \Comment{
Too few particles are merged, the threshold needs to be lowered}

\EndIf
\EndIf
\EndWhile
\Return  $\mathbf{W}_0$ \Comment{Return particle set after merging}
\end{algorithmic}
\end{algorithm}
\end{figure}

\section{Experiment details and more results}
\label{Appendix:exp}

\subsection{Training details} \label{sec:app:args}
The detailed configurations of \texttt{VD-Gen} are listed in Table~\ref{app:table:unet-param} and Table~\ref{app:table:param} \footnote{The codes of the 3D U-net model are implemented based on https://github.com/wolny/pytorch-3dunet}. We did not tune these hyper-parameters for now, a better performance could be achieved with well-tuned hyper-parameters.

\begin{table}[ht]
  \centering
  \small
  \vspace{-10pt}
  \caption{Settings for the 3D U-Net models in \texttt{VD-Gen}.}
  \vskip 0.15in
  \label{app:table:unet-param}
    \begin{tabular}{l|l}
    \toprule
    Name & Value \\
    \hline
    Number of U-Net encoders & 5 \\
    Number of U-Net decoders & 5 \\
    Output channels in each encoder & 16, 32, 64, 128, 256 \\
    Convolution kernel size & 3 \\
    Pooling kernel size & 2 \\
    Batch size & 16 \\
    Max training steps & 500k \\
    Warmup steps & 20K \\
    Peak learning rate & 2e-4 \\
    Adams $\epsilon$ & 1e-6 \\
    Adams($\beta_1$, $\beta_2$) & (0.9,0.99) \\
    Gradient clip norm & 0.5 \\
    \bottomrule
    \end{tabular}
\vskip -0.1in
\end{table}

\begin{table}[ht]
  \centering
  \small
  \vspace{-10pt}
  \caption{Settings for SE(3) models in \texttt{VD-Gen}.} 
  \vskip 0.15in
  \label{app:table:param}
    \begin{tabular}{l|l}
    \toprule
     Name & Value\\
    \hline
    \multicolumn{2}{l}{Training}\\
    \hline
    Particle encoder layers & 12 \\
    Pocket encoder layers & 15 \\
    Particle-Pocket Attention layers & 3 \\
    Peak learning rate & 5e-5 \\
    Batch size & 32 \\
    Max training steps & 100k \\
    Warmup steps & 10K \\
    Attention heads & 64 \\
    FFN dropout & 0.1 \\
    Attention dropout & 0.1  \\
    Embedding dropout & 0.1 \\
    Weight decay & 1e-4 \\
    Embedding dim & 512 \\
    FFN hidden dim & 2048 \\
    Gaussian kernel channels & 128 \\
    Activation function & GELU \\
    Learning rate decay & Linear \\
    Adams $\epsilon$ & 1e-6 \\
    Adams($\beta_1$, $\beta_2$) & (0.9,0.99) \\
    Gradient clip norm & 1.0 \\
    Loss weight of $\mathcal{L}_{an}$ in \textit{Particle Initialization} & 1.0 \\
    
    Loss weight of \textit{Particle Movement} & 1.0\\
    Loss weight of $\mathcal{L}_{Merge}$ in  \textit{Molecule Extraction} & 10 \\
    Loss weight of $\mathcal{L}_{error\_pred}$ in  \textit{Molecule Extraction} & 0.01 \\
    Loss weight of \textit{Molecule Refinement} & 1.0 \\
    Loss weight for \textit{Confidence Prediction} & 0.01 \\
    $\tau$, the clip value for coordinate loss & 2.0 \\
    $\delta$, the threshold for coordinate regularization & 1.0 \\
    $\zeta$, the filtering threshold in \textit{Molecule Extraction} & 2.0 \\
    $R_1$, Iterations in \textit{Particle Movement}   & sampled from [1, 4] \\
    $R_2$, Iterations in \textit{Molecule Refinement}   & sampled from [1, 4] \\
    $k_{vp}$  times of the number of atom, uses in \textit{Particle Initialization}  & sampled from [16.0, 18.0]\\
    \hline
    \multicolumn{2}{l}{Inference}\\
    \hline
    $R_1$, Iterations in \textit{Particle Movement}  & 4 \\
    $R_2$, Iterations in \textit{Molecule Refinement}  & 16 \\
    \bottomrule
    \end{tabular}
\vskip -0.1in
\end{table}

\subsection{Evaluation Mertic}
\label{Appendix:metric}

\begin{itemize}

\item \emph{3D similarity}. 
We use LIGSIFT~\cite{ligsift} to calculate 3D similarity. However, by default, LIGSIFT will align the input molecules before calculating 3D similarity. But we want to evaluate the generated 3D structure directly, to examine the end-to-end performance. Therefore, we \emph{remove the alignment in LIGSIFT}. 
\item \emph{Vina}. We use AutoDock Vina1.2~\cite{noauthor_autodock_nodate} to get Vina score. In particular, the re-docking will be applied. That is, the binding pose and the conformation of the ligand molecule generated by the model will be ignored, and a new binding pose and a new molecular conformation will be re-calculated by AutoDock Vina1.2. We believe the re-docking in Vina cannot reflect the actual performance of the pocket-based 3D molecular generation. But to be consistent with previous works, we still use it as one of the metrics. 
\item \emph{Vina*}. 
Vina* is Vina without re-docking. In particular, we use the built-in energy optimization process based on Vina scoring function in AutoDock Vina1.2~\cite{noauthor_autodock_nodate} to minimize the energy of the binding pose of generated molecules, and then use the Vina scoring function to score the energy-minimized binding pose to get Vina* score.
\item \emph{MM-PBSA}. 
We take the default settings of parameters (i.e., solvation mode: GB-2\cite{GB-2}, protein forcefield: amber03\cite{amber03ff}, ligand charge method: bcc\cite{am1-bcc}, dielectric constant: 4.0) and workflow (i.e., force field building, structure optimization by energy minimization, MM/GB(PB)SA calculation) of ~\cite{yang_wang_zheng_2022} to calculate MM-PBSA score. 
Since the crystal structure indicates the preferred binding pose against a specific target, we filtered the generated molecules by 3D similarity to the molecule in crystal structure and take the molecules whose 3D similarity score is over 0.4 as effective molecules, and we only calculate the MM-PBSA score for the effective molecules.
In Table~\ref{app:tab:pbsa} we show MM-PBSA S.R. (success rate), which calculates the proportion of effective MM-PBSA of the generated molecules. For MM-PBSA B.T. and MM-PBSA Rank we have:
\begin{equation}
\begin{aligned}
\text{MM-PBSA B.T.} =  \frac{1}{n_p}\sum_{i=1}^{n_p} \frac{|\{ g\in\mathcal{G}| \text{MM-PBSA}(g)< \text{MM-PBSA}(\overline{m}_i)\}|}{|\mathcal{G}|},
\end{aligned}
\end{equation}
\begin{align}
\text{MM-PBSA Rank} =  \frac{1}{n_p}\sum_{i=1}^{n_p} \text{rank}_i,
\end{align}
where $n_p$ is the number of proteins in the test set, $\mathcal{G}$ represents the generated molecular set, $\overline{m}_i$ represents the molecular in the crystal structure of the $i$-th protein and $\text{rank}_i$ represents the ranking index of the current model among all of the compared models under the $i$-th protein which is ranked by \text{MM-PBSA}. 

\item Metric for ablation studies. We use 3D similarity between the generated molecules and the ground truth as the metric in ablation studies since it reflects the generative ability based on the pocket structure and there is a strong correlation between 3D similarity and binding affinity according to Table~\ref{tab:binding_overall}.

\end{itemize}

\subsection{More Results} \label{sec:app:res}

In Table~\ref{app:tab:vina}, we report more percentile results for Vina, Vina*. In Table~\ref{app:tab:pbsa}, we report more percentile MM-PBSA results and MM-PBSA S.R. scores.
The MM-PBSA S.R. scores in many baselines are very low. Thus, there are not enough effective MM-PBSA results to calculate percentile results in some baselines. Therefore, in each pocket, we replace the failed MM-PBSA result with the worst one generated by that baseline. And we calculated the percentile results after the replacement.

\begin{table*}[ht]
  \centering
  \small
  \vspace{-10pt}
  \caption{More results on Vina and Vina*.} 
  \vskip 0.15in
  \label{app:tab:vina}
  \addtolength{\tabcolsep}{-2pt}
    \begin{tabular}{c|cc|cc|cc|cc}
    \toprule
    \multirow{2}{*}{Model} & \multicolumn{2}{c|}{5-th} &  \multicolumn{2}{c|}{10-th} &  \multicolumn{2}{c|}{25-th} &  \multicolumn{2}{c}{50-th}\\
     & Vina($\downarrow$) & Vina*($\downarrow$) & Vina($\downarrow$)  & Vina*($\downarrow$)   & Vina($\downarrow$) & Vina*($\downarrow$) & Vina($\downarrow$) & Vina*($\downarrow$) \\
    \midrule
    LiGAN \cite{ragoza_generating_2022} &  -6.724 &  -5.372 & -6.324 &  -4.922 &  -5.740 &  -4.215 &  -5.065 &  -3.49 \\
    3DSBDD \cite{sbdd}  &  -8.662 &  -7.227 &  -8.296 &  -6.664 &  -7.557 &  -5.633 &  -6.474 &  -4.078 \\
    GraphBP \cite{liu_generating_2022} & -8.710 & -3.689 & -7.832 & -2.774  & -6.765 & -1.169 & -5.625 & -1.2 \\
    Pocket2Mol \cite{peng_pocket2mol_2022} & -8.332 &  -6.525 & -8.015 &  -5.399 & -7.467 &  -3.513 &  -6.837 &  -1.808 \\
    \texttt{VD-Gen} & ${\textbf{-8.998}}$ & ${\textbf{ -7.398}}$ & {$\textbf{ -8.569}$} & ${\textbf{ -6.736}}$  & ${\textbf{ -7.892}}$ & ${\textbf{-5.738}}$  & ${\textbf{-7.206}}$ & ${\textbf{ -4.549}}$ \\
    
    \bottomrule
    \end{tabular}
\vskip -0.1in
\end{table*}

\begin{table*}[ht]
  \centering
   \small
  \vspace{-10pt}
  \caption{More MM-PBSA results.} 
  \vskip 0.15in
  \label{app:tab:pbsa}
  \addtolength{\tabcolsep}{-1pt}
    \begin{tabular}{c|c|c|c|c|c}
    \toprule
   \multirow{2}{*}{Model} & 5-th &  10-th &  25-th & 50-th & MM-PBSA- \\
      & MM-PBSA($\downarrow$) & MM-PBSA($\downarrow$) &  MM-PBSA($\downarrow$)   & MM-PBSA ($\downarrow$) & S.R.(\%$\uparrow$)\\
    \midrule
    LiGAN \cite{ragoza_generating_2022}  & -17.865  & -13.374 & -8.775 & -7.418 &  11.9  \\
    3DSBDD \cite{sbdd}   &  -30.221 &   -23.623 &  -13.544 & -7.739 & 12.9 \\
    GraphBP \cite{liu_generating_2022} & -5.130 &  -4.894 &  -4.894 &  -4.894 & 0.2 \\
    Pocket2Mol \cite{peng_pocket2mol_2022} & -7.823 & -5.945  & -5.398 &  -5.398 & 1.8\\
    \texttt{VD-Gen} & ${\textbf{ -50.749}}$  & ${\textbf{-46.285}}$  & ${\textbf{ -38.427}}$  &  ${\textbf{  -28.254}}$ & ${\textbf{46.1}}$ \\
    \bottomrule
    \end{tabular}
\vskip -0.1in
\end{table*}

\subsection{Inference Efficiency}

Experiment results have demonstrated the effectiveness of the proposed \texttt{VD-Gen}, and we also check its efficiency here. In particular, we benchmark the inference speed of generating one molecule for 3DSDBB, GraphBP, Pocket2Mol, and our \texttt{VD-Gen}. The results are summarized the Table~\ref{app:inference:speed}. 3DSBDD is the slowest one, due to the inefficient MCMC sampling. Although GraphBP is the fastest one, its generated molecules are the worst. \texttt{VD-Gen} and Pocket2Mol are similar in efficiency. But \texttt{VD-Gen} significantly outperforms Pocket2Mol in effectiveness. Due to the large number of VPs and several movement iterations, 
it is expected that \texttt{VD-Gen} is not the fastest one. We leave the efficiency improvement to future work.

\begin{table}[H]
\small
  \centering
  \vspace{-5pt}
  \caption{Inference Efficiency.} 
  \vskip 0.15in
  \label{app:inference:speed}
    \begin{tabular}{c|ccccc}
    \toprule
     Model &  3DSBDD & GraphBP & Pocket2Mol &  \texttt{VD-Gen} \\
    \midrule
    Time(s)($\downarrow$) & 14.153 & 1.660 & 3.476 & 3.678\\
    \bottomrule
    \end{tabular}
  \vspace{-5pt}
\vskip -0.1in
\end{table}

\subsection{Molecular optimization task} \label{Appendix:model_optim}

\begin{figure*}[ht]%
 \vspace{-0.1cm}
\includegraphics[width=1.0\linewidth]{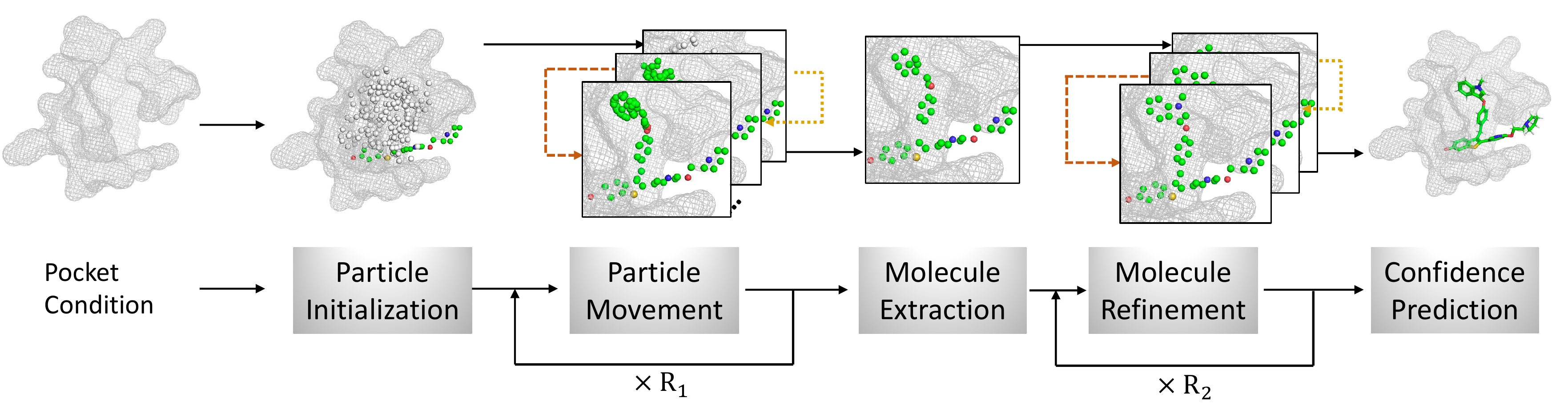}%
\caption{Extending \texttt{VD-Gen} to molecular optimization.} \label{fig:model_optim}
\end{figure*}

\paragraph{Difference in training molecular optimization models} To train the molecular optimization model, we make the following changes. The specific pipeline for molecular optimization task is shown in Fig~\ref{fig:model_optim}.

\begin{itemize}
    \item Rather than predicting the whole molecule, in this task our model is to predict part of the molecule. So  25\% to 40\% atoms are removed from the original molecule and are to predict while others are taken as the input.
    \item During training, the number of VPs is also much smaller, only 8 times of the real atoms.
    \item The VPs are not distributed in the whole pocket cavity, but distributed around the removed atoms.
\end{itemize}

\paragraph{Experiment}
We compare our model with a traditional molecular fragments optimization model DeepFrag~\cite{green2021deepfrag}. DeepFrag can replace molecular fragments based on SMILES, which is a 1D model without pocket information. The results are shown in Table \ref{app:tab:opt:vina} and Table \ref{app:tab:opt:pbsa}. From them, it is clear that \texttt{VD-Gen} can outperform the baseline in molecular optimization.

\begin{table*}[ht]
  \centering
  \small
  \vspace{-10pt}
  \caption{Full percentile results on Vina and Vina*, in molecular optimization tasks.} 
  \vskip 0.15in
  \label{app:tab:opt:vina}
  \addtolength{\tabcolsep}{-2pt}
    \begin{tabular}{c|cc|cc|cc|cc}
    \toprule
    \multirow{2}{*}{Model} & \multicolumn{2}{c|}{5-th} &  \multicolumn{2}{c|}{10-th} &  \multicolumn{2}{c|}{25-th} &  \multicolumn{2}{c}{50-th}\\
     & Vina($\downarrow$) & Vina*($\downarrow$) & Vina($\downarrow$)  & Vina*($\downarrow$)   & Vina($\downarrow$) & Vina*($\downarrow$) & Vina($\downarrow$) & Vina*($\downarrow$) \\
    \midrule
    DeepFrag\cite{green2021deepfrag} &   -8.357 &  - & -8.132 &  - &  -7.775 &  - &  -7.372 &  - \\
    \texttt{VD-Gen} & ${\textbf{-8.868}}$ & {-7.501} & ${\textbf{ -8.595}}$ & {-7.096}  & ${\textbf{-8.150}}$ & { -6.343}  & ${\textbf{ -7.676}}$ & {-5.333} \\
    \bottomrule
    \end{tabular}
  \label{tab:percentlie_optim_vina}
\vskip -0.1in
\end{table*}

\begin{table}[H]
  \centering
  \small
  \caption{Full percentile results on MM-PBSA, in molecular optimization tasks.} 
  \vskip 0.15in
  \label{app:tab:opt:pbsa}
  \addtolength{\tabcolsep}{-2pt}
    \begin{tabular}{c|c|c|c|c|c}
    \toprule
   \multirow{2}{*}{Model} & 5-th &  10-th &  25-th & 50-th\\
      & MM-PBSA($\downarrow$) & MM-PBSA($\downarrow$) &  MM-PBSA($\downarrow$)   & MM-PBSA ($\downarrow$) & MM-PBSA B.T.($\uparrow$) \\
    \midrule
    DeepFrag\cite{green2021deepfrag} & -51.783 & -48.959  & -39.786 & -34.485 & 23.9 \\
    \texttt{VD-Gen} & ${\textbf{ -53.281}}$  & ${\textbf{-51.398}}$  & ${\textbf{-47.152}}$  &  ${\textbf{-39.276}}$ & ${\textbf{30.0}}$  \\
    \bottomrule
    \end{tabular}
  \label{tab:percentlie_optim_pbsa}
\vskip -0.1in
\end{table}

\end{document}